\documentclass[10pt]{article}

\usepackage[utf8]{inputenc}
\usepackage[a4paper,lmargin=1.5in, rmargin=1.5in, footnotesep=1cm plus 4pt minus 4pt]{geometry}
\usepackage{setspace}
\usepackage{amsmath,amsthm,amssymb,esint}
\usepackage{mathtools, nccmath}
\allowdisplaybreaks
\usepackage{multirow}
\usepackage{sansmath}
\usepackage{pgfplots}
\pgfplotsset{compat = newest}
\usepackage{listings}
\usepackage{color}
\usepackage{placeins}
\usepackage{multicol}
\usepackage{multirow}
\usepackage{enumitem}
\usepackage{lipsum}
\usepackage{cases}
\usepackage{epstopdf}
\usepackage{float}
\usepackage{booktabs}
\usepackage{dirtytalk}
\usepackage{subcaption}
\usepackage{graphicx}

\usepackage{comment}

\usepackage[T1]{fontenc}
\usepackage{lmodern}
\usepackage{times}

\usepackage{sectsty}
\allsectionsfont{\sffamily\bfseries}
\subsubsectionfont{\sffamily\mdseries\itshape}

\usepackage[font=sf,labelfont=bf]{caption}

\usepackage[font=sf]{caption}

\newtheoremstyle{mystyle1}
  {}
  {}
  {\itshape}
  {}
  {\sffamily\bfseries}
  {{\sffamily\mdseries .}}
  { }
  {}
\theoremstyle{mystyle1}

\newtheorem{prop}{Proposition}

\newtheoremstyle{mystyle2}
  {}
  {}
  {}
  {}
  {\itshape\sffamily}
  {\textsf{.}}
  { }
  {}
\theoremstyle{mystyle2}
\newtheorem*{prf}{Proof}

\usepackage{hyperref}
\usepackage{xcolor}
\definecolor{rsrs}{RGB}{19, 59, 123}
\definecolor{myred}{rgb}{0.7, 0.0, 0.0}
\definecolor{mygreen}{rgb}{0.0, 0.3, 0.0}
\hypersetup{
    colorlinks,
    linkcolor={rsrs},
    citecolor={rsrs},
    urlcolor={rsrs}
}

\usepackage{tocloft}

\usepackage[nottoc]{tocbibind}

\usepackage{kantlipsum}

\usepackage[
    backend=biber,
    citestyle=authoryear-icomp,
    bibstyle=authoryear,
    natbib=true,
    maxbibnames=99,
    maxcitenames=2,
    giveninits=true,
    terseinits=true,
    uniquelist=false
]{biblatex}
\DefineBibliographyStrings{english}{%
    andothers = {\em et\addabbrvspace al.}
}
\DeclareNameAlias{author}{last-first}
\DeclareFieldFormat*{title}{#1}
\renewbibmacro{in:}{}

\citetrackerfalse
\DefineBibliographyStrings{english}{%
  mathesis = {MSc thesis},
}

\usepackage[title]{appendix}

\usepackage{authblk}

\usepackage[hang]{footmisc}
\usepackage[ruled, algo2e]{algorithm2e} 
\usepackage{algpseudocode}
\SetKw{KwInitialization}{Initialize}
\SetKw{KwInput}{Input:}
\SetKw{KwOutput}{Output:}

\usepackage{changepage}

\usepackage{booktabs}
\title{Exjobb}
\begin{document}

\title{ {\sffamily Direct optimization of dose--volume histogram metrics in radiation therapy treatment planning} }
\author[$*$\textsf{,}$\dagger$]{Tianfang Zhang}
\author[$\dagger$]{Rasmus Bokrantz}
\author[$*$]{Jimmy Olsson}
{
\affil[$*$]{Mathematical Statistics, Department of Mathematics, KTH Royal Institute of Technology, Stockholm SE-100 44, Sweden}
\affil[$\dagger$]{RaySearch Laboratories, Sveavägen 44, Stockholm SE-103 65, Sweden}
}
\date{\textsf{August 27, 2020}}
\maketitle

\begin{quote}
{\centering
\section*{Abstract}
}
We present a method of directly optimizing on deviations in clinical goal values in radiation therapy treatment planning. Using a new mathematical framework in which metrics derived from the dose--volume histogram are regarded as functionals of an auxiliary random variable, we are able to obtain volume-at-dose and dose-at-volume as infinitely differentiable functions of the dose distribution with easily evaluable function values and gradients. Motivated by the connection to risk measures in finance, which is formalized in this framework, we also derive closed-form formulas for mean-tail-dose and demonstrate its capability of reducing extreme dose values in tail distributions. Numerical experiments performed on a prostate and a head-and-neck patient case show that the direct optimization of dose--volume histogram metrics produced marginally better results than or outperformed conventional planning objectives in terms of clinical goal fulfilment, control of low- and high-dose tails of target distributions and general plan quality defined by a pre-specified evaluation measure. The proposed framework eliminates the disconnect between optimization functions and evaluation metrics and may thus reduce the need for repetitive user interaction associated with conventional treatment planning. The method also has the potential of enhancing plan optimization in other settings such as multicriteria optimization and automated treatment planning.
\newline
\begin{spacing}{0.9}
{\sffamily\small \noindent \textbf{Keywords:} Dose--volume histogram, clinical goals, mean-tail-dose, objective functions, smooth approximation, inverse planning.}
\end{spacing}
\end{quote}






\section{Introduction}

Radiation therapy treatment planning can be a time-consuming process, often requiring the planner to perform several re-optimizations before a plan with satisfactory plan quality can be obtained. The criteria used for assessing plan quality are usually communicated through clinical goals, but the actual objectives and constraints of the optimization problem to be solved typically comprise penalty functions not directly related to the clinical goals. Since many clinical goals are specified in terms of dose--volume criteria, which are too cumbersome for most large-scale gradient-based optimization solvers to handle in practice using their exact formulations \citep{ehrgott}, an artificial disconnect is introduced between plan optimization and plan evaluation, dating back to the very advent of the use of mathematical optimization for inverse planning \citep{bortfeldbook}.

Several approaches to reducing the need for repetitive user interaction have been proposed in the literature. One such approach is multicriteria optimization (MCO) \citep{miettinen, rasmus, breedveld}, which enables the articulation of preferences in real time after a set of Pareto optimal plans has been obtained---however, as the tradeoff functions used in MCO are typically the same as for ordinary plan optimization, it does not intrinsically solve the problem of the objective functions of the optimization problem being distinct from the plan evaluation criteria. In recent years, plenty of research has gone into automatic planning methods using machine learning where, in general, data consisting of historically delivered clinical plans is used to estimate suitable parameters of some predetermined optimization problem (see, e.g., \citealt{ge} or \citealt{siddique} for reviews on the subject). Examples include the prediction of weights in a weighted penalty-sum formulation \citep{boutilier} and the learning of appropriate weight adjustment schemes to satisfy clinical goals using deep reinforcement learning \citep{shen}, but also the mimicking of a dose distribution, a dose--volume histogram (DVH) or other dose-related quantities predicted to be achievable for the current patient \citep{mcintosh, appenzoller, ng}. Moreover, a variety of methods \citep{dai, mukherjee} based on the exact mixed-integer programming formulation of optimization under dose--volume constraints, originally described by \citet{langer}, have appeared in the literature. However, although they use different techniques to reduce the large computational burden of solving such a problem, the resulting execution times still make the methods impractical for most clinical settings \citep{ehrgott}.

Other approaches have aimed at developing more exact as well as computationally tractable surrogates for dose--volume criteria than the conventionally used penalty functions \citep{bortfeldbook}. Approximations of volume-at-dose have been proposed in \citet{scherrer} and \citet{liu}, which all rely on the replacement of step functions by sigmoids; \citet{fu} instead approximated the step functions by ramp functions to obtain a convex formulation. \citet{romeijn} first outlined the use of mean-tail-dose criteria in place of dose--volume criteria, which was further investigated by \citet{lovisa} in an MCO formulation. \citet{zarepisheh} demonstrated the possibility of mimicking a set of reference DVHs using moment-based functions, based on a relation  shown in \citet{zinchenko} between appropriately defined equivalent uniform dose (EUD) statistics and the DVH curve. Similarly, \citet{liu} suggested the use of a kurtosis-based metric for controlling extreme values in dose distributions.

In this paper, we present a new perspective on DVH-based metrics based on the suitable definition of an auxiliary random variable. By assuming an independent, nonzero noise of each voxel dose observation, we are able to obtain equivalent formulations of volume-at-dose and dose-at-volume as infinitely differentiable functions with easily evaluable gradients, rendering them tractable for gradient-based optimization in a way which previously has not been possible. Similarly, we provide novel closed-form expressions for the equivalent of mean-tail-dose and its gradient and, furthermore, show that it is a convex function of dose. Analogous formulas are also given for homogeneity index (HI) and conformity index (CI). We demonstrate the advantages of being able to optimize directly on clinical goals by comparing to conventional penalty functions on a prostate and a head-and-neck patient case using volumetric modulated arc therapy (VMAT). In particular, we show the potential of achieving better dose--volume criterion satisfaction using objectives formulated in terms of deviations in clinical goal values, as well as better tail distribution control using mean-tail-dose. 

\section{Methods}
\label{methods}

\subsection{Mathematical notation}

Before proceeding with the construction of DVH-based metrics, we first establish necessary notation and recall some conventional definitions. Given a region of interest (ROI) $R$ in the patient volume, encoded as an index set of voxels, let $d_R = (d_i)_{i \in R}$ be the dose vector containing the dose $d_i$ delivered to each voxel $i \in R$ and let $r_R = (r_i)_{i \in R}$ be the vector containing the respective voxel volumes relative to the volume of $R$, satisfying $\sum_{i \in R} r_i = 1$. For convenience of notation, we shall frequently omit the subscript $R$ when the associated ROI is clear from context. We use $1_A$ for the indicator equaling $1$ when the predicate $A$ is true and $0$ otherwise, $\delta$ for the Kronecker delta $\delta_{ii'} = 1_{i \: = \: i'}$ and $(x)_-$, $(x)_+$ for the negative and positive part functions $-\min\{x, 0\}$ and $\max\{x, 0\}$, respectively.

\subsection{Conventional DVH-based metrics}

\begin{figure}[h]
\centering
\includegraphics[width = 0.7\textwidth]{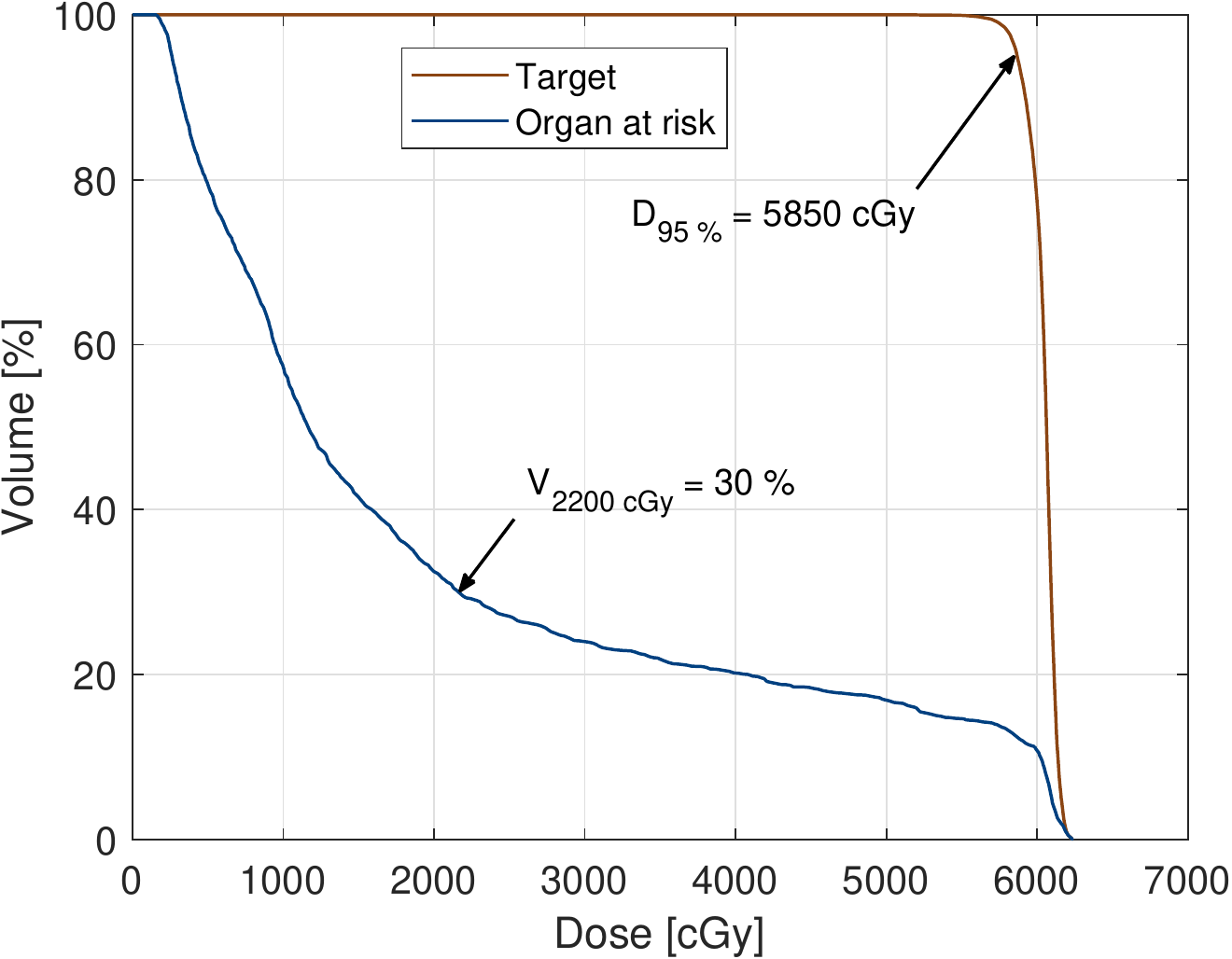}
\caption{Illustration of how the metrics volume-at-dose and dose-at-volume relate to DVH curves.}
\label{dvhsimple}
\end{figure}

In treatment planning, the clinical goals of a plan are often formulated in terms of volume-at-dose and dose-at-volume. The volume-at-dose $\operatorname{V}_x$ at dose level $x$ is a function of the dose vector $d = d_R$, defined as
\[
\operatorname{V}_x(d) = \sum_{i \in R} r_i 1_{d_i \: \geq \: x}
,\]
and the dose-at-volume $\operatorname{D}_v$ at volume level $0 \leq v \leq 1$ is its generalized inverse, defined as
\[
\operatorname{D}_v(d) = \operatorname{inf}\{x \in \mathbb{R} : \operatorname{V}_x(d) \leq v\}
.\]
An at-least dose--volume criterion $(\hat{d}, \hat{v})$ with respect to a reference dose $\hat{d}$ and a reference volume $\hat{v}$ is the goal or requirement that $\operatorname{V}_{\hat{d}}(d) \geq \hat{v}$ or, equivalently, that $\operatorname{D}_{\hat{v}}(d) \geq \hat{d}$; for at-most criteria, the inequalities are reversed. A plot of $\operatorname{V}_x(d)$ against $x$ is recognized as the DVH associated with $d$---see Figure \ref{dvhsimple}.

In their original forms, volume-at-dose and dose-at-volume functions are not suitable for gradient-based optimization as they are discontinuous with respect to dose---instead, it is common to optimize on penalty functions associated with the dose--volume criteria. For $(\hat{d}, \hat{v})$, the so called min-DVH and max-DVH functions $\operatorname{DVH}_{\hat{d}, \hat{v}}^-$, $\operatorname{DVH}_{\hat{d}, \hat{v}}^+$ corresponding to, respectively, the at-least and at-most criteria are given by \citep{bortfeldbook}
\[
\operatorname{DVH}_{\hat{d}, \hat{v}}^-(d) = \int_0^{\hat{v}} \big( \mathrm{D}_v(d) - \hat{d} \big)_-^2 \, dv \quad \text{and} \quad \operatorname{DVH}_{\hat{d}, \hat{v}}^+(d) = \int_{\hat{v}}^1 \big( \mathrm{D}_v(d) - \hat{d} \big)_+^2 \, dv
.\]

Often mentioned drawbacks of dose--volume criteria include the fact that they are nonconvex functions of dose, which can lead to the existence of multiple local optima of the associated optimization problem, and the fact that they offer limited control of tail values. To address these problems, \citet{romeijn} proposed the use of mean-tail-dose functions as a surrogate for dose--volume criteria. In particular, the authors relied on an indirect formulation by \citet{rockafellaruryasev} in which the value of mean-tail-dose was written as the global optimum of an optimization problem, which was later exploited further by \citet{lovisa}. As the formulation requires the introduction of additional variables and constraints to the original optimization problem, however, which significantly increases computation times, mean-tail-dose functions defined in such a way are impractical in a general-purpose optimization framework.

\subsection{Smooth DVH-based metrics}

It is often stated \citep{lovisa, liu} that dose-at-volume and mean-tail-dose have counterparts in finance in terms of value-at-risk and conditional value-at-risk \citep{rockafellaruryasev, hlhr}. Since risk measures are typically defined as functionals of random variables \citep{hlhr} and since many of the criteria commonly used for plan quality evaluation depend on the dose $d$ solely through the DVH, we introduce the auxiliary random variable $D$ intended to capture the distributional characteristics of $d$. More precisely, letting $I$ be the random variable such that $\mathbb{P}(I = i) = r_i$ for each $i \in R$ and $\varepsilon = \epsilon Z$, where $Z$ is a standard normal random variable and $\epsilon \geq 0$ is a constant, we define $D$ as 
\[
D = d_I + \varepsilon
,\]
where $I$ and $\varepsilon$ are assumed to be independent. Here, it is understood that all random variables are defined on a common probability space $(\Omega, \mathcal{F}, \mathbb{P})$. The variable $\varepsilon$ can be interpreted as an observation noise added upon $d_I$---its purpose will become apparent later.

Letting $\epsilon = 0$ for now, we note that volume-at-dose can be reformulated as $\operatorname{V}_x(d) = \mathbb{P}(D \geq x)$ and dose-at-volume as $\mathrm{D}_v(d) = \inf\{ x \in \mathbb{R} : \mathbb{P}(D \geq x) \leq v \}$. Thus, $\operatorname{D}_v$ is equal to the $v$-level value-at-risk using $D$ as the discounted portfolio loss, and $\operatorname{MTD}^+_v$ is the $v$-level conditional value-at-risk \citep{hlhr}. In general, any scalar-valued function $\psi$ of dose $d$ may be written as
\[
\psi(d) = \Psi(D)
,\]
where $\Psi$ is a functional on some space of random variables.\footnote{Note that this is possible since $d$ can be recovered from $D : \Omega \to \mathbb{R}$ as $d = (d_{I(\omega_i)})_{i \in R} = (D(\omega_i))_{i \in R}$ when $\epsilon = 0$, each $\omega_i \in \Omega$ being such that $I(\omega_i) = i$.} Functions $\psi$ such that $\Psi$ only depends on $D$ through its distribution will be called DVH-based, capturing the notion that some functions only depend on the dose through the corresponding DVH.

\subsubsection{Volume-at-dose and dose-at-volume}

Note that some of the definitions given above are, in fact, unnecessarily awkward in order to handle the fact that $D$ almost surely takes values in a discrete set when $\epsilon = 0$. Therefore, let now $\epsilon > 0$ so that $D$ has a density supported on the whole real line---we shall see that this leads to an explicit formula for the gradient of dose-at-volume. Denoting by $k$ and $K$ the probability density function and the cumulative distribution function, respectively, of $\varepsilon$, given by
\[
k(x) = \frac{1}{\epsilon \sqrt{2\pi}} \operatorname{exp}\!\left( -\frac{x^2}{2\epsilon^2} \right) \quad \text{and} \quad K(x) = \frac{1}{2} \operatorname{erfc}\!\left( -\frac{x}{\epsilon\sqrt{2}} \right)
,\]
we can write 
\[
\operatorname{V}_x(d) = \mathbb{P}(D \geq x) = \operatorname{\mathbb{E}} \mathbb{P}(d_I + \varepsilon \geq x \mid I) = \sum_{i \in R} r_i K(d_i - x)
.\]
As $\operatorname{V}_x(d)$ is now everywhere strictly decreasing in $x$, $\operatorname{D}_v(d)$ becomes uniquely defined by the implicit relation 
\begin{equation}
\label{dav}
v = \sum_{i \in R} r_i K(d_i - \operatorname{D}_v(d)),
\end{equation}
which also leads to the following:
\begin{prop}[Gradient of dose-at-volume]
\label{davgradprop}
For $0 < v < 1$ and $\epsilon > 0$, $\operatorname{D}_v$ is a continuously differentiable function with gradient given componentwise by
\begin{equation}
\label{davgrad}
\frac{\partial \mathrm{D}_v(d)}{\partial d_i} = \frac{r_i k(d_i - \operatorname{D}_v(d))}{\sum_{i' \in R} r_{i'} k(d_{i'} - \mathrm{D}_v(d))}
\end{equation}
for each $i \in R$.
\end{prop}
\begin{prf}
Differentiation of (\ref{dav}) with respect to $d_i$ yields
\begin{equation*}
\begin{split}
0 &= \frac{\partial}{\partial d_i} \sum_{i' \in R} r_{i'} K(d_{i'} - \operatorname{D}_v(d)) \\
&= \sum_{i' \in R} r_{i'} k(d_{i'} - \operatorname{D}_v(d)) \!\left( \delta_{ii'} - \frac{\partial \mathrm{D}_v(d)}{\partial d_i} \right) \\
&= r_i k(d_i - \operatorname{D}_v(d)) - \frac{\partial \mathrm{D}_v(d)}{\partial d_i} \sum_{i' \in R} r_{i'} k(d_{i'} - \operatorname{D}_v(d)),
\end{split}
\end{equation*}
and the result follows upon rearrangement and the fact that the denominator in (\ref{davgrad}) is always positive. \qed
\end{prf}
\noindent In fact, a stronger result than once continuous differentiability is possible (a proof is given in Appendix \ref{davinfdiff}):
\begin{prop}[Infinite differentiability of dose-at-volume]
\label{infdiffprop}
For $0 < v < 1$ and $\epsilon > 0$, $\operatorname{D}_v$ is infinitely differentiable.
\end{prop}

Hence, with above facts established, the function value and gradient of dose-at-volume are straightforward to evaluate given any dose $d$, e.g. in the following way:
\begin{enumerate}
\item find $\operatorname{D}_v(d)$ from the relation (\ref{dav}) using any numerical root-finding algorithm, and
\item evaluate the gradient using Proposition \ref{davgradprop}.
\end{enumerate}
Naturally, this significant advantage comes at a cost of having to introduce a noise $\varepsilon$, but provided that its standard deviation is small relative to the doses, the noisy dose-at-volume will differ relatively little from its noise-free counterpart, as will be showcased in Section \ref{hncase}. The form of $\operatorname{V}_x$ coincides with that used in \citet{scherrer} and \citet{liu} but with $K$ as the sigmoid approximation of the step function---in fact, this is essential since Proposition \ref{closedformulasmtd} relies on Proposition \ref{davgradprop} and the fact that $\varepsilon$ is normally distributed. Interestingly, one can also note that Bayes' theorem gives the interpretation
\[
\frac{\partial \mathrm{D}_v(d)}{\partial d_i} = \mathbb{P}(I = i \mid D = \mathrm{D}_v(d))
\]
for all $i \in R$, which means that voxels with dose relatively closer to $\operatorname{D}_v(d)$ will have relatively larger contribution to the gradient. Since $\sum_{i \in R} \partial \mathrm{D}_v(d) / \partial d_i = 1$, the gradient will never vanish. 

\subsubsection{Mean-tail-dose}

We can proceed by deriving similar formulas for mean-tail-dose functions, which we also show are convex, making them particularly well-suited for gradient-based optimization. In particular, when the dose vector is linear in the decision variables (such as in fluence map optimization; \citealt{ehrgott}), exclusively using mean-tail-dose functions for objectives and constraints will lead to a convex optimization problem. The lower and upper mean-tail-doses $\operatorname{MTD}^-_{v}$ and $\operatorname{MTD}^+_{v}$ at volume level $v$ can be defined as
\[
\operatorname{MTD}^-_v(d) = \mathbb{E}[D \mid D \leq \operatorname{D}_v(d)] \quad \text{and} \quad \operatorname{MTD}^+_v(d) = \mathbb{E}[D \mid D \geq \operatorname{D}_v(d)]
\]
for $D$ with everywhere positive density. Our results are summarized in the following and proven in Appendix \ref{mtdderivations}:
\begin{prop}[Closed formulas for mean-tail-dose]
\label{closedformulasmtd}
For $0 < v < 1$ and $\epsilon > 0$, $\operatorname{MTD}^-_{v}$ and $\operatorname{MTD}^+_{v}$ are obtained as
\[
\operatorname{MTD}^-_v(d) = \frac{1}{1 - v} \sum_{i \in R} r_i \Big( d_i K(\mathrm{D}_v(d) - d_i) - \epsilon^2 k(\mathrm{D}_v(d) - d_i) \Big)
\]
and
\[
\operatorname{MTD}^+_v(d) = \frac{1}{v} \sum_{i \in R} r_i \Big( d_i K(d_i - \mathrm{D}_v(d)) + \epsilon^2 k(d_i - \mathrm{D}_v(d))\Big)
,\]
and their dose gradients are given by the relations
\[
\frac{\partial \mathrm{MTD}^-_v(d)}{\partial d_i} = \frac{1}{1 - v} r_i K(\mathrm{D}_v(d) - d_i) \quad \text{and} \quad \frac{\partial \mathrm{MTD}^+_v(d)}{\partial d_i} = \frac{1}{v} r_i K(d_i - \mathrm{D}_v(d))
.\]
Moreover, $-\operatorname{MTD}^-_v$ and $\operatorname{MTD}^+_v$ are convex functions.
\end{prop}

\subsubsection{Average dose}

The corresponding version of the average dose function $\operatorname{EUD}_1$, which is a special case of EUD \citep{zinchenko}, is given by
\[
\operatorname{EUD}_1(d) = \operatorname{\mathbb{E}} D = \operatorname{\mathbb{E}} d_I = \sum_{i \in R} r_i d_i
\]
as $\operatorname{\mathbb{E}} \varepsilon = 0$. This coincides with the conventional formulation.

\subsubsection{Homogeneity index and conformity index}

Accordingly, gradients for other well-known DVH-based metrics, which in their conventional definitions are nondifferentiable, can be derived in similar fashions. For instance, using Proposition \ref{davgradprop}, the homogeneity index $\operatorname{HI}_v$ at volume level $v$ \citep{hi}, which is here defined for $1/2 \leq v \leq 1$ as 
\[
\operatorname{HI}_v(d) = \frac{\operatorname{D}_{v}(d)}{\operatorname{D}_{1 - v}(d)}
,\]
has dose derivative 
\[
\frac{\partial \mathrm{HI}_v(d)}{\partial d_i} = \frac{r_i}{\operatorname{D}_{1-v}(d)^2} \!\left( \frac{\mathrm{D}_{1 - v}(d) k(d_i - \mathrm{D}_v(d))}{\sum_{i' \in R} r_{i'} k(d_{i'} - \mathrm{D}_v(d))} - \frac{\mathrm{D}_v(d) k(d_i - \mathrm{D}_{1 - v}(d))}{\sum_{i' \in R} r_{i'} k(d_{i'} - \mathrm{D}_{1 - v}(d))} \right)
\]
for each $i \in R$. Similarly, given the external ROI $E$ (that is, the full treatment volume) and a target ROI $T \subseteq E$, the conformity index $\operatorname{CI}_x$ at isodose level $x$ with respect to $T$ and $E$ \citep{ci} is here defined as
\[
\operatorname{CI}_x(d) = \alpha \frac{\mathrm{V}_x(d_{T})}{\mathrm{V}_x(d_E)}
,\]
where $\alpha$ is the absolute volume ratio between $T$ and $E$. Denoting by $r_T$ and $r_E$ the local relative voxel volumes, its dose derivative is then obtained as
\[
\frac{\partial \mathrm{CI}_x(d)}{\partial d_i} = \alpha \frac{k(x - d_i)}{\operatorname{V}_x(d_E)^2} \Big( 1_{i \: \in \: T} r_{T, i} \mathrm{V}_x(d_E) - r_{E, i} \mathrm{V}_x(d_T) \Big)
.\]

\subsection{Optimization formulation}
\label{optimizationformulation}

Following the possibilities of optimizing directly on DVH-based metrics according to the above, we can construct objective functions and constraints directly corresponding to requirements of the metrics. We will assume that the overall plan quality can be judged by a pre-specified plan quality metric and compare direct optimization of this with what may be achieved using conventional planning objectives. Such plan quality metrics were first introduced by \citet{nelms} and have been further developed by, for instance, \citet{cilla}, and their direct optimization has been investigated by \citet{bjorn}.

Although one could use any plan quality metric in principle, we shall choose a relatively simplistic form for our numerical experiments to better highlight properties of functions derived from the proposed framework. In particular, given a set $\{\psi_j\}_{j \in O \cup C}$ of DVH-based metrics and goals on the form $\psi_j(d) \leq \hat{\psi}_j$ or $\psi_j(d) \geq \hat{\psi}_j$, each specified as either an objective ($j \in O$) or a constraint ($j \in C$), we use a loss function $L_j$ to measure the loss $L_j(\psi_j(d), \hat{\psi}_j)$ associated with achieving the value $\psi_j(d)$ relative to the acceptance value $\hat{\psi}_j$. We use the partial weighted sum
\[
L_{O}(d) = \sum_{j \in O} \frac{w_j}{\hat{\psi}_j} L_j(\psi_j(d), \hat{\psi}_j)
\]
as our loss contribution due to the objectives, where $w_j$ represents the importance weight for the (relative) loss of each $\psi_j$, along with losses on the form of ramp functions
\[
L^-(x, \hat{x}) = (x - \hat{x})_- \quad \text{and} \quad L^+(x, \hat{x}) = (x - \hat{x})_+
.\]
Moreover, to account for the effect of eventual constraint infeasibilities, we use
\[
L_C(d) = \sum_{j \in C} \frac{w_C^2}{\hat{\psi}_j^2} L_j(\psi_j(d), \hat{\psi}_j)^2
\]
for the corresponding loss contribution. The reason for using squared ramp functions is to better reflect the notion that small infeasibilities often are acceptable whereas larger infeasibilities often are not. Note that the constraints share the same weight $w_C$, which can be set to a relatively high value. The overall plan quality metric $L_{\mathrm{tot}}$ is then given by $L_{\mathrm{tot}}(d) = L_O(d) + L_C(d)$.

Hence, letting $\eta$ be the optimization variables with feasible set $\mathcal{E}$, from which the total dose $d$ can be determined by some dose deposition mapping $d = d(\eta)$ (see, for example, \citet{vmat} for details on this), the optimization problem for the case of direct optimization of clinical goals can be written simply as 
\begin{equation*}
\begin{aligned}
\label{newformulation}
& \underset{\eta \in \mathcal{E}}{\text{minimize}}
& & L_{\mathrm{tot}}(d(\eta)).
\end{aligned}
\end{equation*}
This is compared to a conventional formulation using quadratic-penalty functions, written as
\begin{equation*}
\begin{aligned}
\label{conventionalformulation}
& \underset{\eta \in \mathcal{E}}{\text{minimize}}
& & \sum_{j \in O} \frac{w_j^2}{\hat{\psi}_j^2} \operatorname{DVH}_j^{\pm}(d(\eta)) + \sum_{j \in C} \frac{\widetilde{w}_C^2}{\hat{\psi}_j^2} \operatorname{DVH}_j^{\pm}(d(\eta))
\end{aligned}
\end{equation*}
each min-DVH or max-DVH function $\operatorname{DVH}_j^{\pm}$ matching the reference dose and reference volume of the corresponding dose--volume criterion---for average-dose criteria, we instead use the associated penalty functions; HI and CI criteria are ignored. The squaring of the weights are intended to compensate for the fact that the penalty functions are quadratic whereas our loss functions are linear in dose units. Note also that we use a separate constraint weight $\widetilde{w}_C$, which need not be equal to $w_C$, to ensure that the extent to which constraints are preserved is comparable to that of the direct plan quality metric optimization.

\subsection{Computational study}

The above methods were tested numerically by the authors on a prostate and a head-and-neck patient treatment case using VMAT delivered by an Elekta Agility treatment system (Elekta, Stockholm, Sweden). The formulations in Section \ref{newformulation} were implemented in a development version of RayStation 10A (RaySearch Laboratories, Stockholm, Sweden). The respective optimization problems were solved using RayStation's native sequential quadratic programming solver, where each optimization comprised the following stages: 50 iterations of fluence map optimization, conversion to machine parameters, and between three to five runs of direct machine parameter optimization, each with 100 iterations and zero optimality tolerance followed by accurate dose calculation, until convergence was reached. Approximate doses during optimization were calculated by a singular value decomposition algorithm \citep{svd} and accurate doses by a collapsed cone algorithm \citep{cc}. For all cases, a uniform dose grid with a voxel resolution of $3 \; \mathrm{mm}$ was used. The smoothness parameter $\epsilon$ was set to $5 \; \mathrm{cGy}$, and $\mathrm{D}_v(d)$ was found from (\ref{dav}) using a Newton's method search. 

For the prostate case, we used a single $360$-degree arc with control points spaced $2$ degrees apart. The plan quality metric was set up according to Table \ref{prostateclingoals}. Both the CI goal on the planning target volume (PTV) and the dose--volume goal on the external ROI were used to control plan conformity. To show the potential of mean-tail-dose controlling tails in dose distributions, we also tried replacing the goals on $\operatorname{D}_{2 \, \%}$ and $\operatorname{D}_{98 \, \%}$ in the PTV by $\operatorname{MTD}_{2 \, \%}^+$ and $\operatorname{MTD}_{98 \, \%}^{-}$, respectively, using the same acceptance levels. The squared constraint weight $w_C^2$ and its counterpart $\widetilde{w}_C^2$ for the conventional formulation were both set to $10^4$. 

\begin{table}[h]
\caption{The constituent clinical goals and weights of the plan quality metric considered for the numerical experiments on the prostate case. The squared constraint weight $w_C^2$ was set to $10^4$.}
\vspace{-0.5cm}
{\bgroup
\def\arraystretch{1.15}
\begin{center}
\begin{tabular}{llll}
ROI & Goal & Weight & Constraint \\ \hline
Prostate & $\operatorname{D}_{99 \, \%} \geq 6000 \; \mathrm{cGy}$ & -- & Yes \\
PTV & $\operatorname{D}_{2 \, \%} \leq 6200 \; \mathrm{cGy}$ & $10$ & No \\
PTV & $\operatorname{D}_{95 \, \%} \geq 5850 \; \mathrm{cGy}$ & $10$ & No \\
PTV & $\operatorname{D}_{98 \, \%} \geq 5700 \; \mathrm{cGy}$ & $10$ & No \\
PTV & $\operatorname{HI}_{95 \, \%} \geq 0.95$ & $10$ & No \\
PTV & $\operatorname{CI}_{6000 \, \mathrm{cGy}} \geq 0.98$ & $10$ & No \\
External & $\operatorname{D}_{2 \, \%} \leq 3000 \; \mathrm{cGy}$ & $5$ & No \\
Rectum wall & $\operatorname{D}_{30 \, \%} \leq 2250 \; \mathrm{cGy}$ & $3$ & No \\
Rectum wall & $\operatorname{D}_{50 \, \%} \leq 1250 \; \mathrm{cGy}$ & $3$ & No \\
Bladder wall & $\operatorname{D}_{25 \, \%} \leq 2000 \; \mathrm{cGy}$ & $3$ & No \\
Left femur & $\operatorname{D}_{5 \, \%} \leq 3000 \; \mathrm{cGy}$ & $1$ & No \\
Right femur & $\operatorname{D}_{5 \, \%} \leq 3000 \; \mathrm{cGy}$ & $1$ & No \\
\end{tabular}
\end{center}
\egroup
}
\label{prostateclingoals}
\end{table}

For the head-and-neck case, two tests were run: one with all goals formulated as weighted objectives, excluding goals on the pharyngeal constrictor muscles (PCMs), and one with most goals formulated as constraints, using only goals on the PCMs as objectives. Both tests used two $360$-degree arcs with control points spaced $2$ degrees apart. The plan quality metrics for the former and latter tests were set up according to Tables \ref{hnclingoals} and \ref{hnclingoals2}, respectively. In the unconstrained formulation, for all dose--volume goals in the targets (but not in the subtraction of the high-dose target from the low-dose target) and for those with relative reference volume less than $1 \; \%$, mean-tail-dose was used instead of dose-at-volume in the direct clinical goal optimization due to their tail-controlling abilities showcased in the prostate case. In the mostly constrained formulation, however, to ensure a fair comparison with equally restricting constraints, no goals were replaced by mean-tail-dose. The optimizations started in the solution obtained from the direct, unconstrained optimization which had already satisfied all the constraint goals (see Section \ref{hncase}). Here, $w_C^2$ was set to $10^4$ while we tried $10^3$, $10^4$, $10^5$ and $10^6$ for $\widetilde{w}_C^2$ in comparison.

\begin{table}[h]
\caption{The constituent clinical goals and weights of the plan quality metric considered for the numerical experiments on the head-and-neck case using the unconstrained formulation.}
\vspace{-0.5cm}
{\bgroup
\def\arraystretch{1.15}
\begin{center}
\begin{tabular}{llll}
ROI & Goal & Weight & Constraint \\ \hline
PTV 7000 & $\operatorname{D}_{98 \, \%} \geq 6650 \; \mathrm{cGy}$ & $10$ & No \\
PTV 7000 & $\operatorname{EUD}_1 \geq 6950 \; \mathrm{cGy}$ & $5$ & No \\
PTV 7000 & $\operatorname{D}_{5 \, \%} \leq 7400 \; \mathrm{cGy}$ & $5$ & No \\
PTV 5425 & $\operatorname{D}_{98 \, \%} \geq 5150 \; \mathrm{cGy}$ & $10$ & No \\
$\text{PTV 5425} \setminus \text{PTV 7000}$ & $\operatorname{D}_{5 \, \%} \leq 5800 \; \mathrm{cGy}$ & $5$ & No \\
External & $\operatorname{D}_{10 \, \%} \leq 3500 \; \mathrm{cGy}$ & $5$ & No \\
Spinal cord & $\operatorname{D}_{0.1 \, \mathrm{cm}^3} \leq 4500 \; \mathrm{cGy}$ & $10$ & No \\
Left parotid & $\operatorname{EUD}_1 \leq 2600 \; \mathrm{cGy}$ & $5$ & No \\
Right parotid & $\operatorname{EUD}_1 \leq 2600 \; \mathrm{cGy}$ & $3$ & No \\
Left submandibular gland & $\operatorname{EUD}_1 \leq 4000 \; \mathrm{cGy}$ & $3$ & No \\
Brain & $\operatorname{D}_{0.1 \, \mathrm{cm}^3} \leq 5000 \; \mathrm{cGy}$ & $1$ & No \\
Brainstem & $\operatorname{D}_{0.1 \, \mathrm{cm}^3} \leq 5600 \; \mathrm{cGy}$ & $1$ & No \\
Anterior left eye & $\operatorname{D}_{0.1 \, \mathrm{cm}^3} \leq 500 \; \mathrm{cGy}$ & $1$ & No \\
Anterior right eye & $\operatorname{D}_{0.1 \, \mathrm{cm}^3} \leq 500 \; \mathrm{cGy}$ & $1$ & No \\
Posterior left eye & $\operatorname{D}_{0.1 \, \mathrm{cm}^3} \leq 500 \; \mathrm{cGy}$ & $1$ & No \\
Posterior right eye & $\operatorname{D}_{0.1 \, \mathrm{cm}^3} \leq 500 \; \mathrm{cGy}$ & $1$ & No \\
\end{tabular}
\end{center}
\egroup
}
\label{hnclingoals}
\end{table}

\begin{table}[h]
\caption{The constituent clinical goals and weights of the plan quality metric considered for the numerical experiments on the head-and-neck case using the mostly constrained formulation. The squared constraint weight $w_C^2$ was set to $10^4$.}
\vspace{-0.5cm}
{\bgroup
\def\arraystretch{1.15}
\begin{center}
\begin{tabular}{llll}
ROI & Goal & Weight & Constraint \\ \hline
PTV 7000 & $\operatorname{D}_{98 \, \%} \geq 6650 \; \mathrm{cGy}$ & -- & Yes \\
PTV 7000 & $\operatorname{EUD}_1 \geq 6950 \; \mathrm{cGy}$ & -- & Yes \\
PTV 7000 & $\operatorname{D}_{5 \, \%} \leq 7400 \; \mathrm{cGy}$ & -- & Yes \\
PTV 5425 & $\operatorname{D}_{98 \, \%} \geq 5150 \; \mathrm{cGy}$ & -- & Yes \\
$\text{PTV 5425} \setminus \text{PTV 7000}$ & $\operatorname{D}_{5 \, \%} \leq 5800 \; \mathrm{cGy}$ & -- & Yes \\
External & $\operatorname{D}_{10 \, \%} \leq 3500 \; \mathrm{cGy}$ & -- & Yes \\
Spinal cord & $\operatorname{D}_{0.1 \, \mathrm{cm}^3} \leq 4500 \; \mathrm{cGy}$ & -- & Yes \\
Left parotid & $\operatorname{EUD}_1 \leq 2600 \; \mathrm{cGy}$ & -- & Yes \\
Right parotid & $\operatorname{EUD}_1 \leq 2600 \; \mathrm{cGy}$ & -- & Yes \\
Left submandibular gland & $\operatorname{EUD}_1 \leq 4000 \; \mathrm{cGy}$ & -- & Yes \\
Brain & $\operatorname{D}_{0.1 \, \mathrm{cm}^3} \leq 5000 \; \mathrm{cGy}$ & -- & Yes \\
Brainstem & $\operatorname{D}_{0.1 \, \mathrm{cm}^3} \leq 5600 \; \mathrm{cGy}$ & -- & Yes \\
Anterior left eye & $\operatorname{D}_{0.1 \, \mathrm{cm}^3} \leq 500 \; \mathrm{cGy}$ & -- & Yes \\
Anterior right eye & $\operatorname{D}_{0.1 \, \mathrm{cm}^3} \leq 500 \; \mathrm{cGy}$ & -- & Yes \\
Posterior left eye & $\operatorname{D}_{0.1 \, \mathrm{cm}^3} \leq 500 \; \mathrm{cGy}$ & -- & Yes \\
Posterior right eye & $\operatorname{D}_{0.1 \, \mathrm{cm}^3} \leq 500 \; \mathrm{cGy}$ & -- & Yes \\
Inferior PCM & $\operatorname{EUD}_{1} \leq 5000 \; \mathrm{cGy}$ & $1$ & No \\
Middle PCM & $\operatorname{EUD}_{1} \leq 5000 \; \mathrm{cGy}$ & $1$ & No \\
Superior PCM & $\operatorname{EUD}_{1} \leq 5000 \; \mathrm{cGy}$ & $1$ & No \\
\end{tabular}
\end{center}
\egroup
}
\label{hnclingoals2}
\end{table}

\section{Results}
\label{results}

We refer to the optimizations with conventional objectives, with direct optimization of DVH-based metrics and with direct optimization of DVH-based metrics and some dose--volume goals replaced by corresponding mean-tail-dose goals as Conv, Direct 1 and Direct 2, respectively. Table \ref{lossresults} shows the objective loss $L_O$, the constraint loss $L_C$ and the overall plan quality metric $L_{\mathrm{tot}}$ for the different optimization formulations on the three test cases. One can observe that for the prostate case and the unconstrained head-and-neck case, all clinical goals were fulfilled for the direct formulations while some clinical goals were left slightly unfulfilled for the conventional formulation. For the mostly constrained head-and-neck case, it was apparent that the direct formulation was superior to all runs using conventional penalty functions in terms of the specified plan quality metric, with $L_O$ increasing and $L_C$ decreasing for increasing $\widetilde{w}_C$.

\begin{table}[h]
\caption{Resulting loss values for the three test cases, compared between the conventional formulation with different $\widetilde{w}_C$ and the direct optimizations of clinical goals, with and without replacing certain dose--volume goals with mean-tail-dose. For head-and-neck, only the version of the direct optimization using mean-tail-dose goals was run for the unconstrained problem, and only that using dose-at-volume goals was run for the mostly constrained problem. Note that definitions of the losses differ between the test cases.}
\vspace{-0.5cm}
{\bgroup
\def\arraystretch{1.15}
\begin{center}
\begin{tabular}{l|l|lll}
 &  & $L_O$ & $L_C$ & $L_{\mathrm{tot}}$ \\ \hline
\multirow{3}{*}{Prostate} & Conv, $\widetilde{w}_C^2 = 10^3$ & $1.0556$ & $0.0025$ & $1.0580$ \\
 & Direct 1 & $0.0000$ & $0.0000$ & $0.0000$ \\
 & Direct 2 & $0.0000$ & $0.0000$ & $0.0000$ \\ \hline
\multirow{2}{*}{HN, unconstrained} & Conv & $0.0599$ & -- & $0.0599$ \\
 & Direct 2 & $0.0000$ & -- & $0.0000$ \\ \hline
\multirow{5}{*}{HN, mostly constrained} & Conv, $\widetilde{w}_C^2 = 10^3$ & $0.1432$ & $1.7412$ & $1.8844$ \\
 & Conv, $\widetilde{w}_C^2 = 10^4$ & $0.3118$ & $0.1787$ & $0.4905$ \\
 & Conv, $\widetilde{w}_C^2 = 10^5$ & $0.3664$ & $0.0171$ & $0.3835$ \\
 & Conv, $\widetilde{w}_C^2 = 10^6$ & $0.4456$ & $0.0000$ & $0.4456$ \\
 & Direct 1 & $0.1834$ & $0.0012$ & $0.1846$ \\ 
\end{tabular}
\end{center}
\egroup
}
\label{lossresults}
\end{table}

\subsection{Prostate case}

Table \ref{prostateresults} shows the particular clinical goal levels after optimization. Eight of the twelve clinical goals were not fulfilled for the conventional formulation, including the constraint for the prostate, although the deviations for the dose--volume goals were relatively small and negligible in practice; on the other hand, all goals were fulfilled for the direct formulations. In particular, it is apparent that the HI and CI goals were not taken into account in the optimization, leading to the relatively large deviation for especially the CI goal. 

\begin{figure}[h]
\centering
\begin{subfigure}[t]{0.5\textwidth}
\centering
\includegraphics[width = \textwidth]{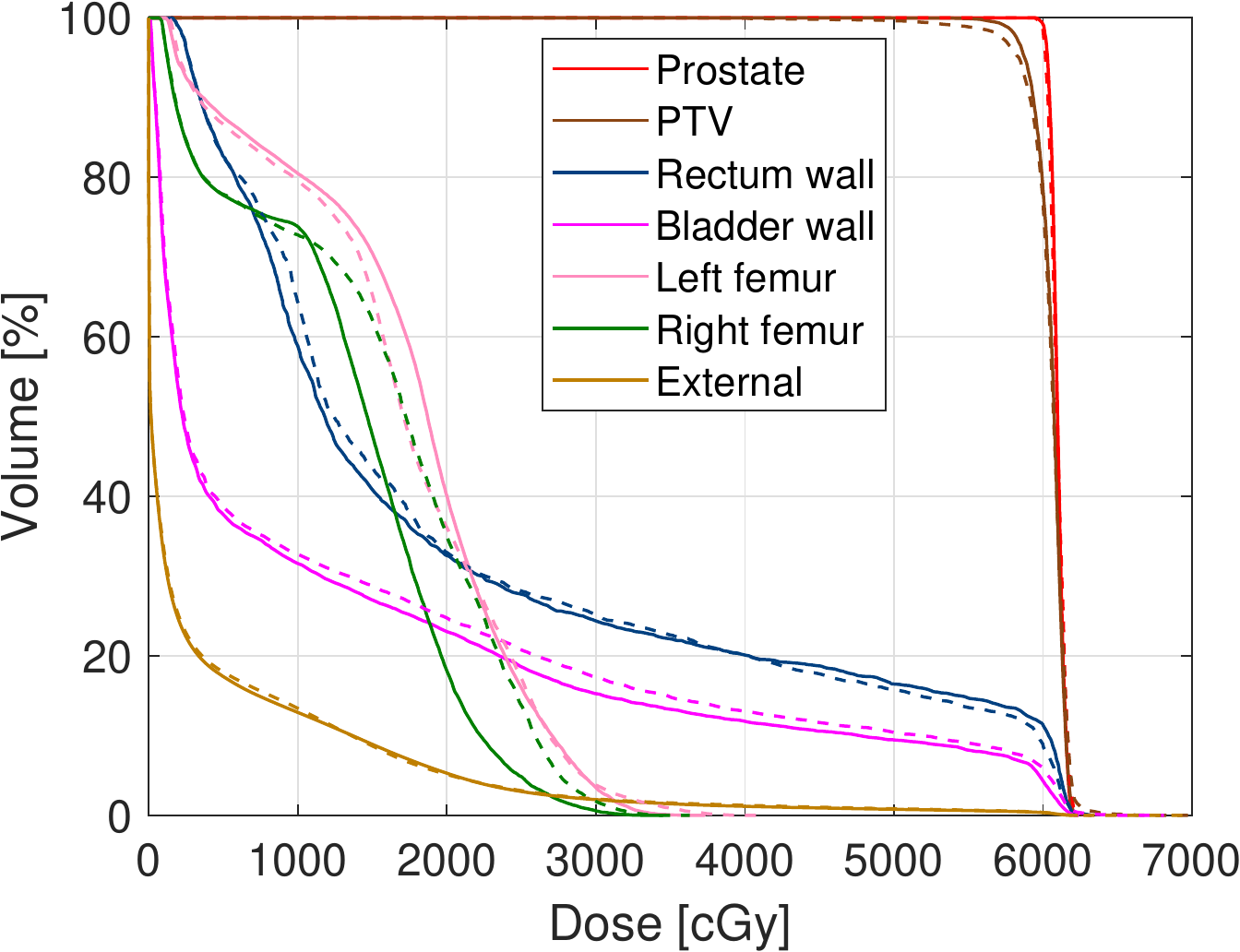}
\caption{}
\end{subfigure}%
~ 
\begin{subfigure}[t]{0.5\textwidth}
\centering
\includegraphics[width = \textwidth]{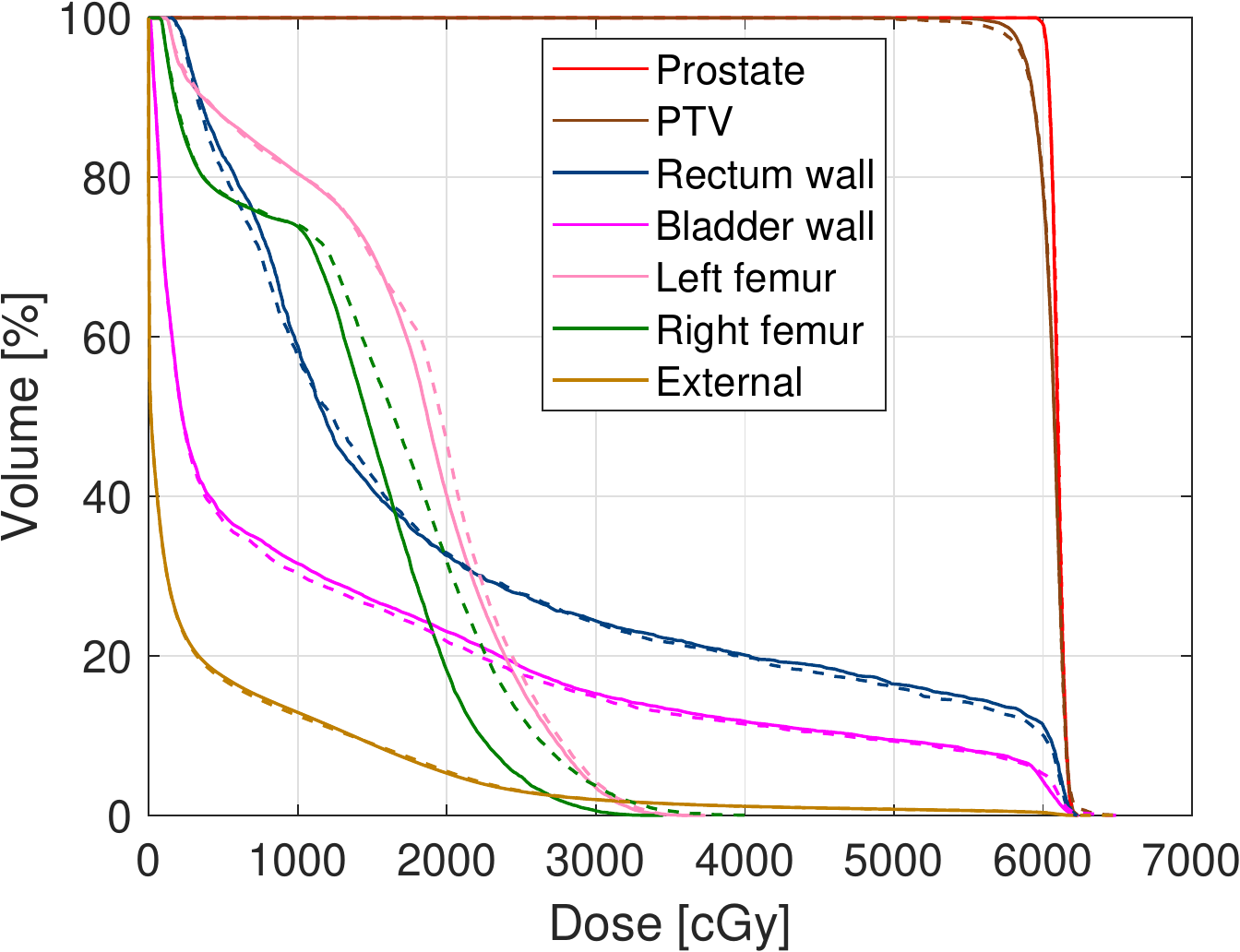}
\caption{}
\end{subfigure}
\caption{DVH comparisons between the optimized plans for the prostate case. (a) shows the direct optimization of clinical goals with mean-tail-dose (solid) compared to the conventional formulation (dashed), and (b) shows the direct optimization of clinical goals with mean-tail-dose (solid) compared to with dose-at-volume (dashed). The $\operatorname{D}_{99.5 \, \%}$--$\operatorname{D}_{0.5 \, \%}$ range in the PTV was $5690 \; \mathrm{cGy}$--$6207 \; \mathrm{cGy}$ and $5565 \; \mathrm{cGy}$--$6222 \; \mathrm{cGy}$ for the direct optimizations with and without mean-tail-dose, respectively, and $5279 \; \mathrm{cGy}$--$6273 \; \mathrm{cGy}$ for the conventional formulation.}
\label{prostatedvhs}
\end{figure}

\begin{figure}[h]
\centering
\begin{subfigure}[t]{0.5\textwidth}
\centering
\includegraphics[width = \textwidth]{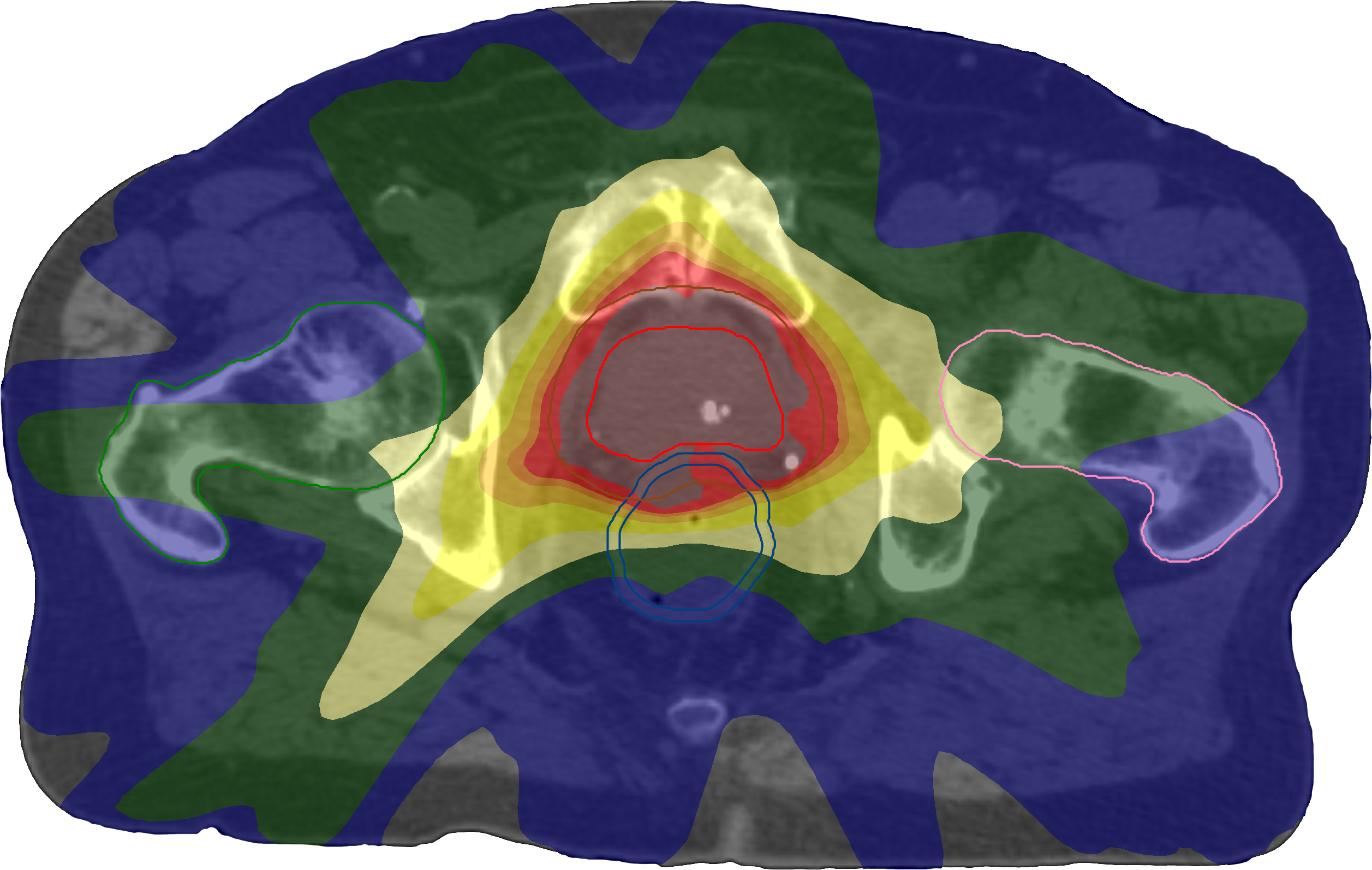}
\caption{}
\end{subfigure}%
~ 
\begin{subfigure}[t]{0.5\textwidth}
\centering
\includegraphics[width = \textwidth]{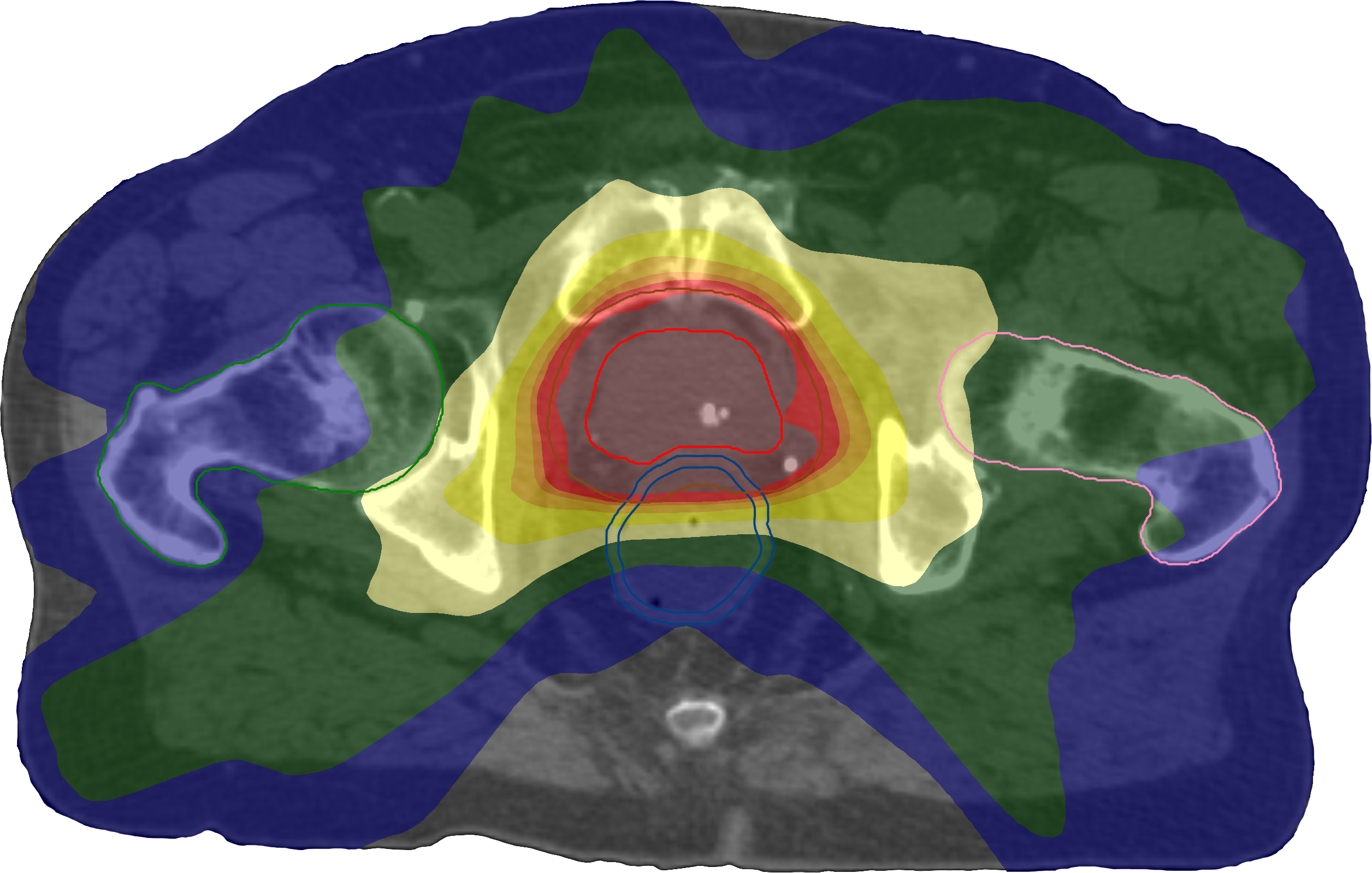}
\caption{}
\end{subfigure}\\[1ex]
\begin{subfigure}[t]{0.7\textwidth}
\centering
\includegraphics[width = \textwidth]{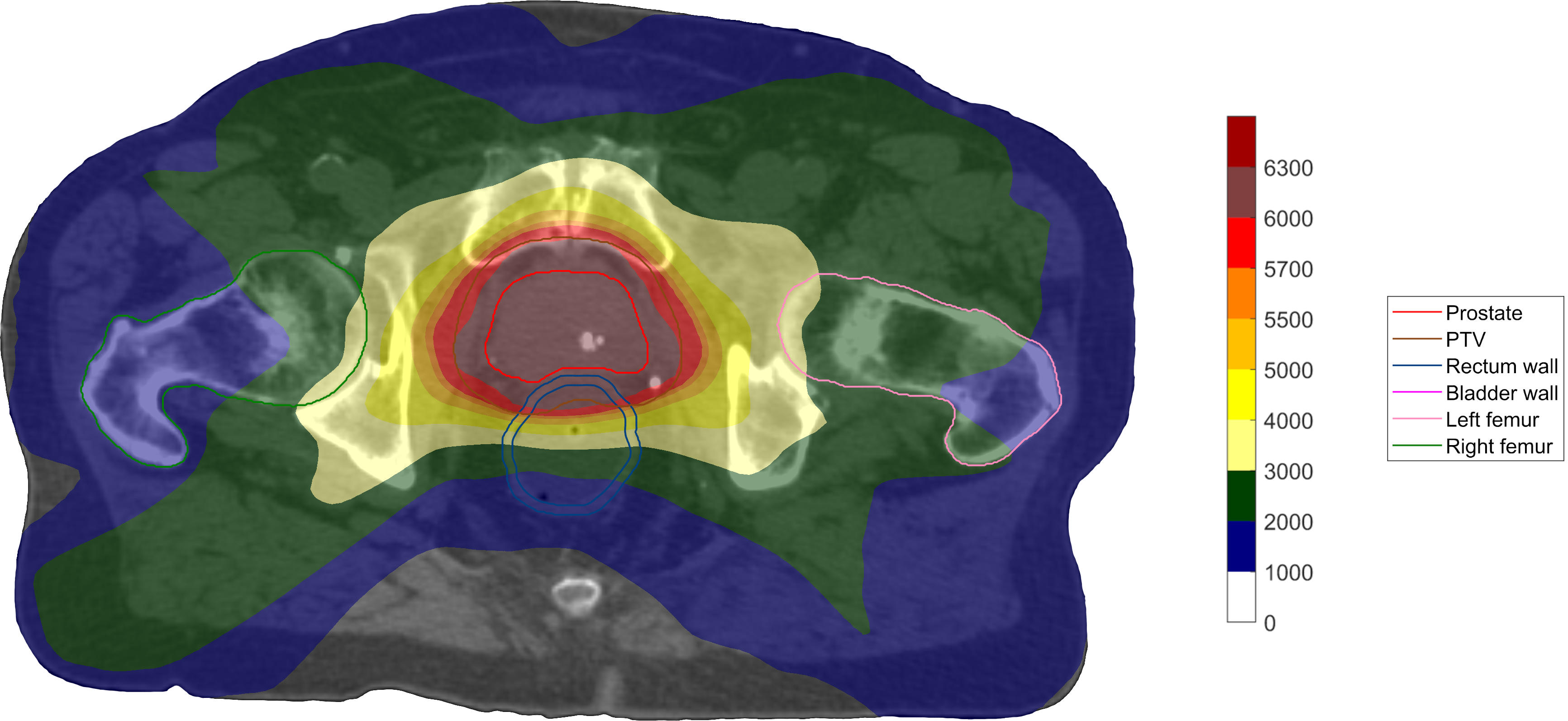}
\caption{}
\end{subfigure}
\caption{Transversal cuts of the spatial dose distributions of the optimized plans on the prostate case using the conventional formulation (a) and the direct formulations with (b) and without (c) replacement of certain dose--volume goals by mean-tail-dose.}
\label{prostatespatial}
\end{figure}

Figure \ref{prostatedvhs} shows DVH comparisons between the formulations, and Figure \ref{prostatespatial} shows the spatial dose distributions. One can observe that the use of mean-tail-dose for the PTV tails leads to fewer extreme values in the lower tail, as can be expected from its properties---the $\operatorname{D}_{99.5 \, \%}$--$\operatorname{D}_{0.5 \, \%}$ range for the PTV was $5690 \; \mathrm{cGy}$--$6207 \; \mathrm{cGy}$ for this case, compared to $5565 \; \mathrm{cGy}$--$6222 \; \mathrm{cGy}$ for the case with only dose-at-volume and $5279 \; \mathrm{cGy}$--$6273 \; \mathrm{cGy}$ for the conventional formulation. It is also possible to see the effect of the goals being made more restrictive when using the same reference volume level in the replacement. Moreover, the relatively large CI shortfall of the conventional plan can be seen by comparing its spatial dose distribution to those of the direct formulations, which actually optimized on the goal. 

\begin{table}[h]
\caption{Resulting clinical goal values for the prostate case, compared between the conventional formulation and the direct optimizations of clinical goals, with and without replacing certain dose--volume goals with mean-tail-dose. Not fulfilled goals are in red and fulfilled goals are in green.}
\vspace{-0.5cm}
{\bgroup
\def\arraystretch{1.15}
\begin{center}
\begin{tabular}{lllll}
ROI & Goal & Conv & Direct 1 & Direct 2 \\ \hline
Prostate & $\operatorname{D}_{99 \, \%} \geq 6000 \; \mathrm{cGy}$ & {\color{myred} $5997 \; \mathrm{cGy}$} & {\color{mygreen} $6003 \; \mathrm{cGy}$} & {\color{mygreen} $6005 \; \mathrm{cGy}$} \\
External & $\operatorname{D}_{2 \, \%} \leq 3000 \; \mathrm{cGy}$ & {\color{myred} $3046 \; \mathrm{cGy}$} & {\color{mygreen} $2990 \; \mathrm{cGy}$} & {\color{mygreen} $2983 \; \mathrm{cGy}$} \\
PTV & $\operatorname{D}_{2 \, \%} \leq 6200 \; \mathrm{cGy}$ & {\color{myred} $6211 \; \mathrm{cGy}$} & {\color{mygreen} $6196 \; \mathrm{cGy}$} & {\color{mygreen} $6185 \; \mathrm{cGy}$} \\
PTV & $\operatorname{D}_{95 \, \%} \geq 5850 \; \mathrm{cGy}$ & {\color{myred} $5838 \; \mathrm{cGy}$} & {\color{mygreen} $5866 \; \mathrm{cGy}$} & {\color{mygreen} $5876 \; \mathrm{cGy}$} \\
PTV & $\operatorname{D}_{98 \, \%} \geq 5700 \; \mathrm{cGy}$ & {\color{myred} $5696 \; \mathrm{cGy}$} & {\color{mygreen} $5742 \; \mathrm{cGy}$} & {\color{mygreen} $5806 \; \mathrm{cGy}$} \\
PTV & $\operatorname{HI}_{95 \, \%} \geq 0.95$ & {\color{myred} $0.94$} & {\color{mygreen} $0.95$} & {\color{mygreen} $0.95$} \\
PTV & $\operatorname{CI}_{6000 \, \mathrm{cGy}} \geq 0.98$ & {\color{myred} $0.90$} & {\color{mygreen} $0.98$} & {\color{mygreen} $0.98$} \\
Rectum wall & $\operatorname{D}_{30 \, \%} \leq 2250 \; \mathrm{cGy}$ & {\color{myred} $2259 \; \mathrm{cGy}$} & {\color{mygreen} $2232 \; \mathrm{cGy}$} & {\color{mygreen} $2240 \; \mathrm{cGy}$} \\
Rectum wall & $\operatorname{D}_{50 \, \%} \leq 1250 \; \mathrm{cGy}$ & {\color{mygreen} $1243 \; \mathrm{cGy}$} & {\color{mygreen} $1230 \; \mathrm{cGy}$} & {\color{mygreen} $1181 \; \mathrm{cGy}$} \\
Bladder wall & $\operatorname{D}_{25 \, \%} \leq 2000 \; \mathrm{cGy}$ & {\color{mygreen} $1970 \; \mathrm{cGy}$} & {\color{mygreen} $1656 \; \mathrm{cGy}$} & {\color{mygreen} $1770 \; \mathrm{cGy}$} \\
Left femur & $\operatorname{D}_{5 \, \%} \leq 3000 \; \mathrm{cGy}$ & {\color{mygreen} $2913 \; \mathrm{cGy}$} & {\color{mygreen} $2959 \; \mathrm{cGy}$} & {\color{mygreen} $2916 \; \mathrm{cGy}$} \\
Right femur & $\operatorname{D}_{5 \, \%} \leq 3000 \; \mathrm{cGy}$ & {\color{mygreen} $2757 \; \mathrm{cGy}$} & {\color{mygreen} $2891 \; \mathrm{cGy}$} & {\color{mygreen} $2483 \; \mathrm{cGy}$} \\
\end{tabular}
\end{center}
\egroup
}
\label{prostateresults}
\end{table}

\subsection{Head-and-neck case}
\label{hncase}

\begin{figure}[h]
\centering
\begin{subfigure}[t]{0.61\textwidth}
\centering
\includegraphics[width = \textwidth]{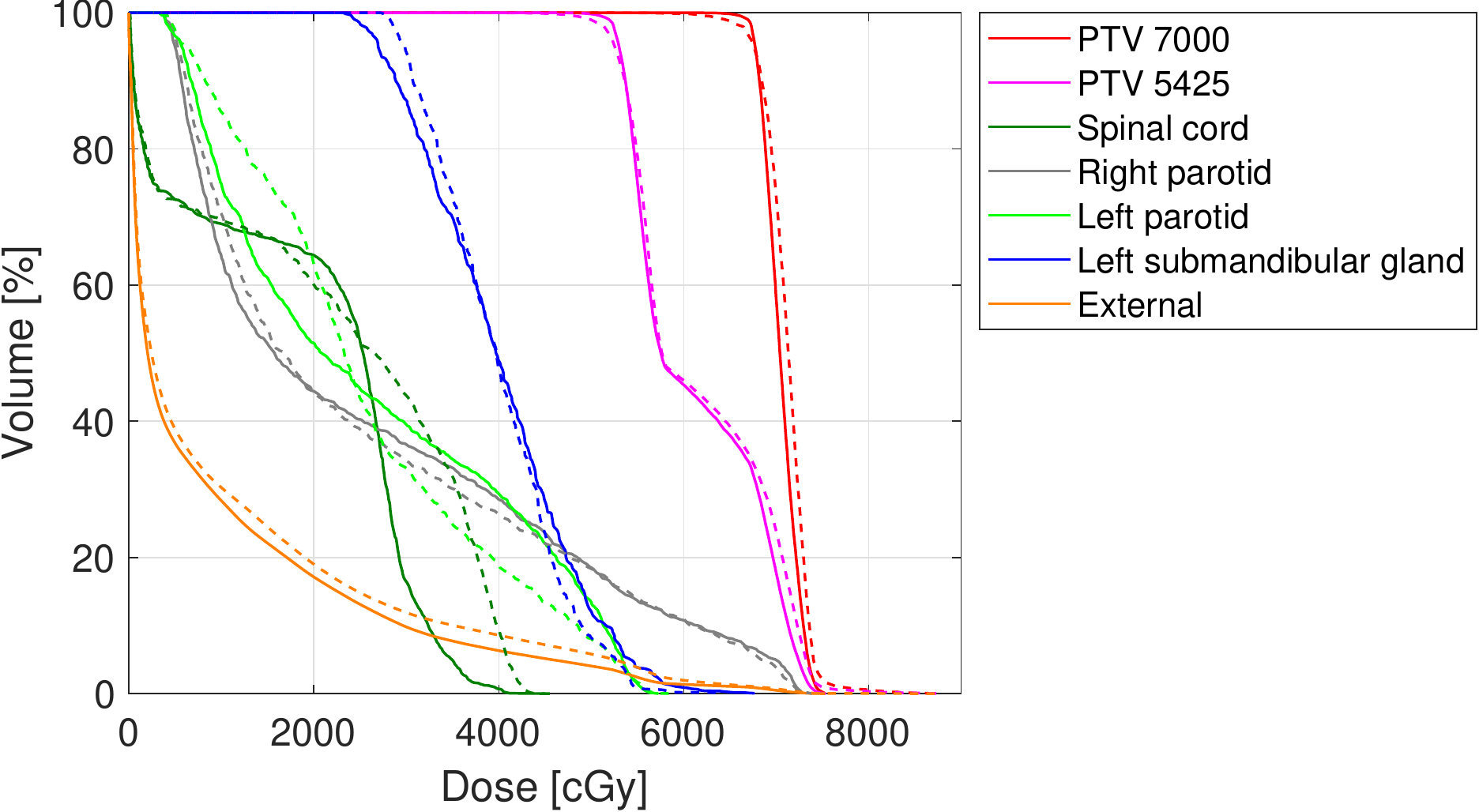}
\caption{}
\label{hndvh}
\end{subfigure}%
~
\begin{subfigure}[t]{0.35\textwidth}
\centering
\includegraphics[width = \textwidth]{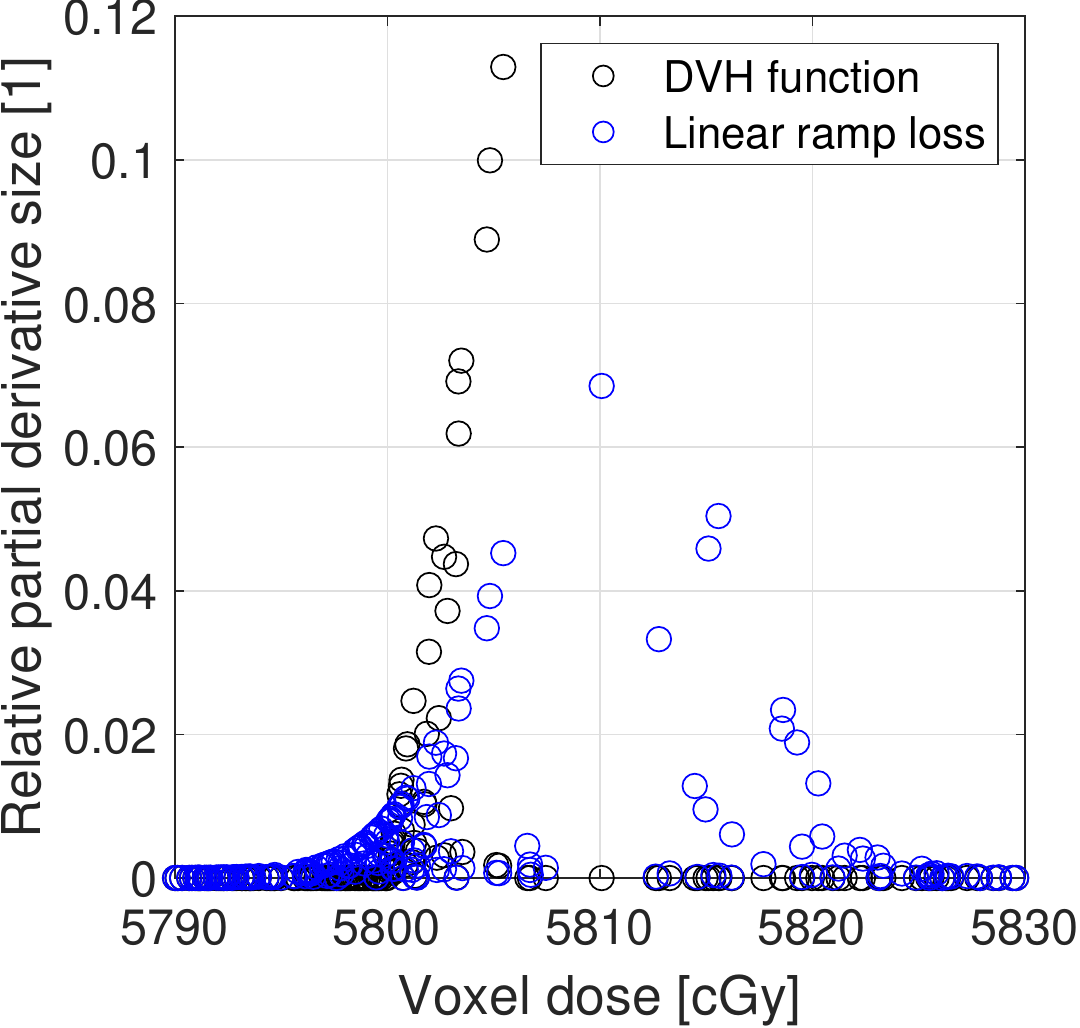}
\caption{}
\label{hngradsizes}
\end{subfigure}
\caption{(a) DVH comparison between the optimized plans obtained for the unconstrained head-and-neck test using the direct optimization of clinical goals (solid) and using conventional functions (dashed). The $\operatorname{D}_{99.5 \, \%}$--$\operatorname{D}_{0.5 \, \%}$ range in the high-dose target was $6620 \; \mathrm{cGy}$--$7463 \; \mathrm{cGy}$ for the former case and $6267 \; \mathrm{cGy}$--$8029 \; \mathrm{cGy}$ for the latter case. (b) Comparison of individual dose derivatives between the quadratic-penalty function and the linear-ramp loss for the dose--volume goal $\operatorname{D}_{0.05} \leq 5800 \; \mathrm{cGy}$ in $\text{PTV 7000} \setminus \text{PTV 5425}$ on the plan obtained using the conventional formulation, where the gradients are normalized to sum to unity.}
\end{figure}

\begin{figure}[h]
\centering
\begin{subfigure}[t]{0.31\textwidth}
\centering
\includegraphics[width = \textwidth]{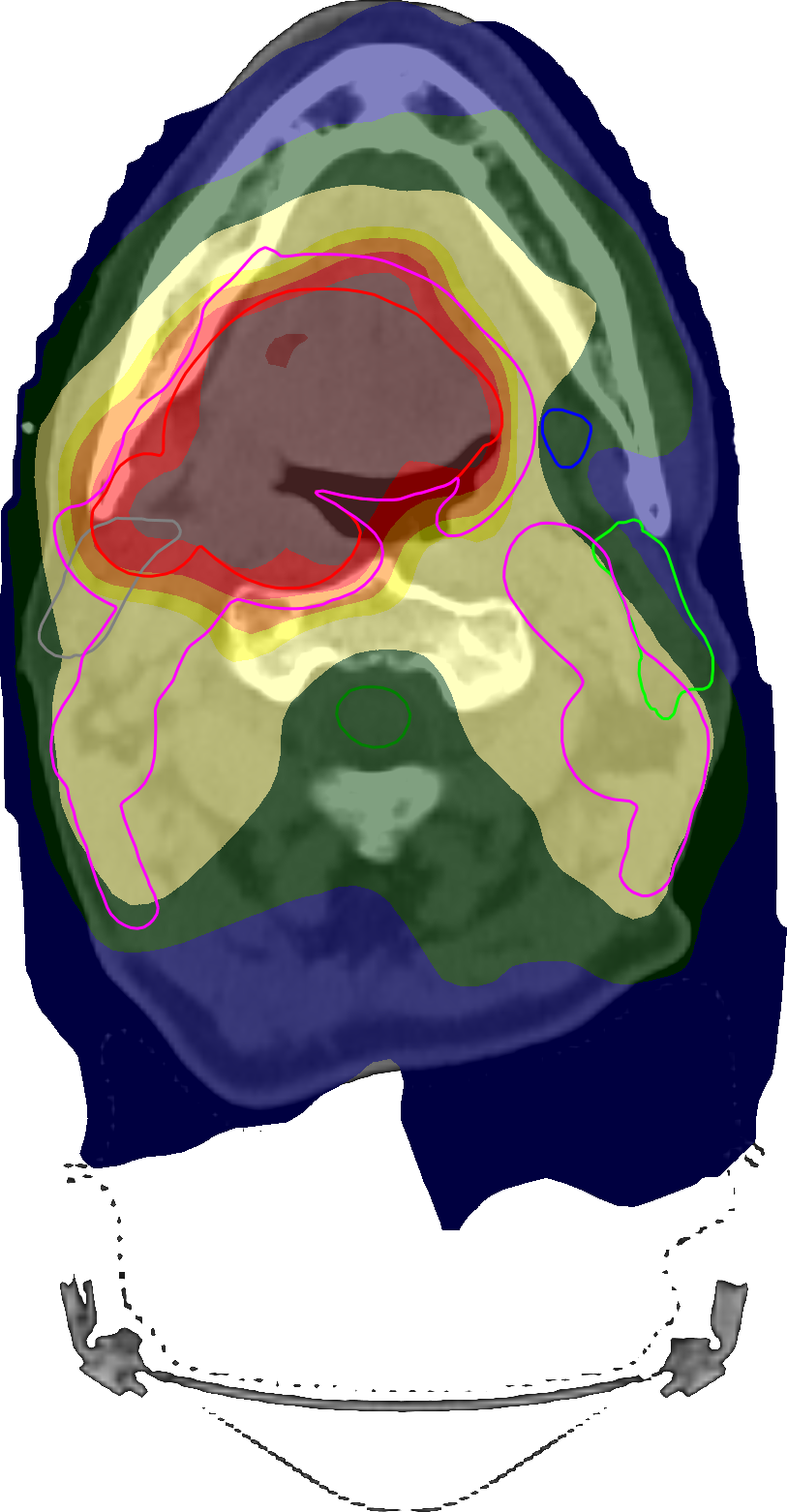}
\caption{}
\end{subfigure}%
\hspace{0.5cm}
\begin{subfigure}[t]{0.54\textwidth}
\centering
\includegraphics[width = \textwidth]{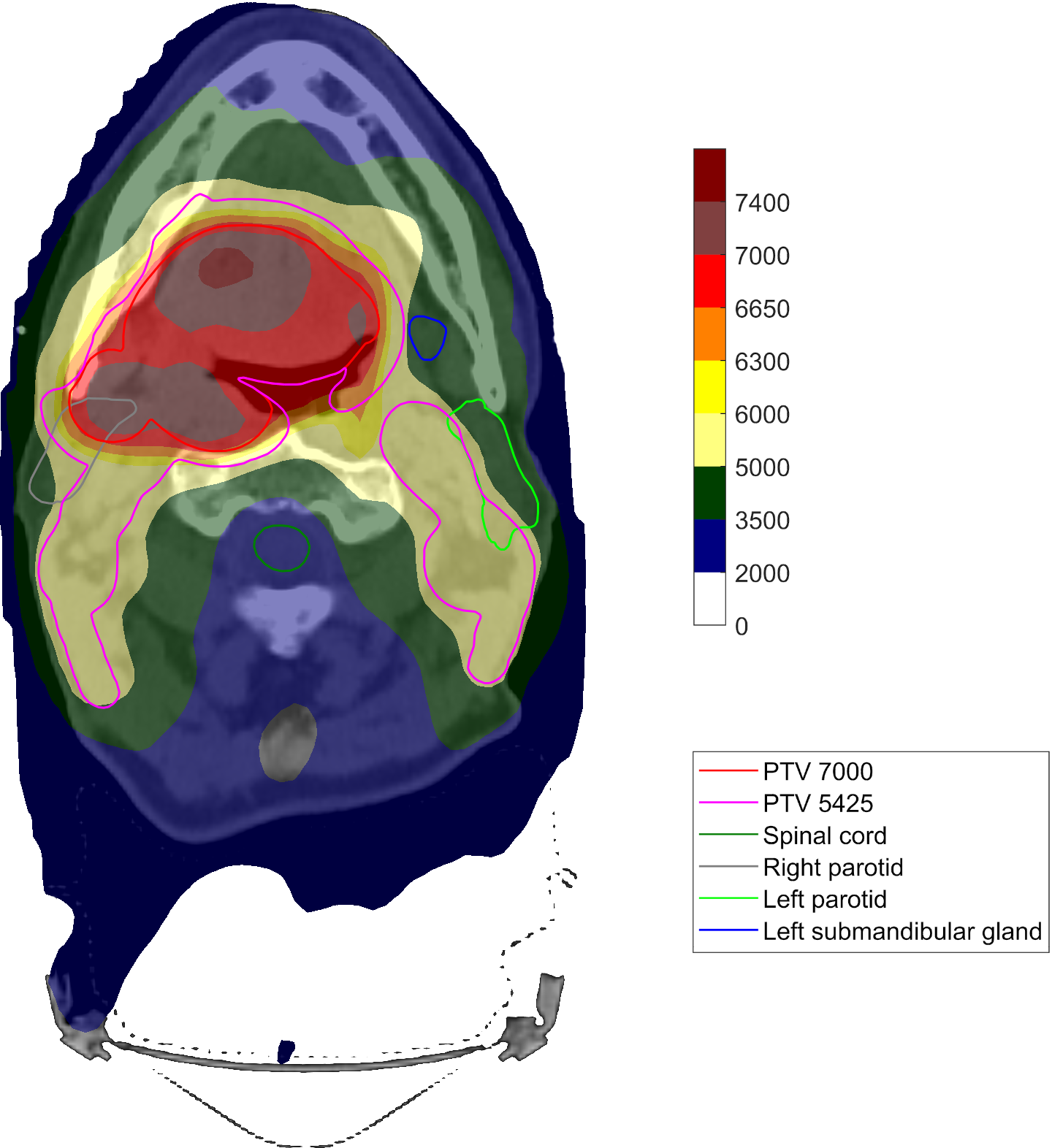}
\caption{}
\end{subfigure}
\caption{Transversal cuts of the spatial dose distributions of the optimized plans obtained for the unconstrained head-and-neck test using the conventional formulation (a) and using the direct optimization of clinical goals (b).}
\label{hnspatial}
\end{figure}

For the unconstrained formulation, Table \ref{hnclingoalsresults} shows the particular clinical goal levels after optimization, and Figures \ref{hndvh} and \ref{hnspatial} show, respectively, the DVHs and the spatial dose distributions. Again, we were able to fulfil all clinical goals with the direct optimization of clinical goals (using only the version including mean-tail-dose this time), whereas the conventional formulation left five clinical goals slightly unfulfilled. In particular, due to the mean-tail-dose functions, both the upper and lower tails of the targets had remarkably fewer extreme values with a $\operatorname{D}_{99.5 \, \%}$--$\operatorname{D}_{0.5 \, \%}$ range in the high-dose target of $6620 \; \mathrm{cGy}$--$7463 \; \mathrm{cGy}$ compared to $6267 \; \mathrm{cGy}$--$8029 \; \mathrm{cGy}$ for the conventional formulation---in fact, in a clinical setting, the latter values would likely have been unacceptable. Unnecessary dose in the external ROI was also significantly reduced, which can be seen both in DVH and in spatial dose. The fact that many goals finished within $1 \; \mathrm{cGy}$ of their acceptance levels for the direct formulation indicates that our choice of noise level leads to an approximation error of dose-at-volume negligible for most purposes. Furthermore, the relatively poor convergence properties of conventional penalty functions can be explained by the fact that their gradients vanish when the underlying clinical goal approaches fulfilment \citep{albin}, whereas this is not the case for the direct formulation. Figure \ref{hngradsizes} shows that smooth dose-at-volume with a ramp loss function leads to more voxels having non-negligible partial derivative than the corresponding conventional penalty function. 

\begin{table}[h]
\caption{Resulting clinical goal values for the unconstrained head-and-neck case, compared between the conventional formulation and the direct optimization of clinical goals using mean-tail-dose. Not fulfilled goals are in red and fulfilled goals are in green.}
\vspace{-0.5cm}
{\bgroup
\def\arraystretch{1.15}
\begin{center}
\begin{tabular}{llll}
ROI & Goal & Conv & Direct 2 \\ \hline
PTV 7000 & $\operatorname{D}_{98 \, \%} \geq 6650 \; \mathrm{cGy}$ & {\color{myred} $6633 \; \mathrm{cGy}$} & {\color{mygreen} $6731 \; \mathrm{cGy}$} \\
PTV 7000 & $\operatorname{EUD}_1 \geq 6950 \; \mathrm{cGy}$ & {\color{mygreen} $7117 \; \mathrm{cGy}$} & {\color{mygreen} $7048 \; \mathrm{cGy}$} \\
PTV 7000 & $\operatorname{D}_{5 \, \%} \leq 7400 \; \mathrm{cGy}$ & {\color{myred} $7402 \; \mathrm{cGy}$} & {\color{mygreen} $7346 \; \mathrm{cGy}$} \\
PTV 5425 & $\operatorname{D}_{98 \, \%} \geq 5150 \; \mathrm{cGy}$ & {\color{myred} $5137 \; \mathrm{cGy}$} & {\color{mygreen} $5242 \; \mathrm{cGy}$} \\
$\text{PTV 5425} \setminus \text{PTV 7000}$ & $\operatorname{D}_{5 \, \%} \leq 5800 \; \mathrm{cGy}$ & {\color{myred} $5809 \; \mathrm{cGy}$} & {\color{mygreen} $5800 \; \mathrm{cGy}$} \\
External & $\operatorname{D}_{10 \, \%} \leq 3500 \; \mathrm{cGy}$ & {\color{myred} $3505 \; \mathrm{cGy}$} & {\color{mygreen} $2962 \; \mathrm{cGy}$} \\
Spinal cord & $\operatorname{D}_{0.1 \, \mathrm{cm}^3} \leq 4500 \; \mathrm{cGy}$ & {\color{mygreen} $4300 \; \mathrm{cGy}$} & {\color{mygreen} $4039 \; \mathrm{cGy}$} \\
Left parotid & $\operatorname{EUD}_1 \leq 2600 \; \mathrm{cGy}$ & {\color{mygreen} $2577 \; \mathrm{cGy}$} & {\color{mygreen} $2597 \; \mathrm{cGy}$} \\
Right parotid & $\operatorname{EUD}_1 \leq 2600 \; \mathrm{cGy}$ & {\color{mygreen} $2593 \; \mathrm{cGy}$} & {\color{mygreen} $2600 \; \mathrm{cGy}$} \\
Left submandibular gland & $\operatorname{EUD}_1 \leq 4000 \; \mathrm{cGy}$ & {\color{mygreen} $3993 \; \mathrm{cGy}$} & {\color{mygreen} $4000 \; \mathrm{cGy}$} \\
Brain & $\operatorname{D}_{0.1 \, \mathrm{cm}^3} \leq 5000 \; \mathrm{cGy}$ & {\color{mygreen} $3798 \; \mathrm{cGy}$} & {\color{mygreen} $2476 \; \mathrm{cGy}$} \\
Brainstem & $\operatorname{D}_{0.1 \, \mathrm{cm}^3} \leq 5600 \; \mathrm{cGy}$ & {\color{mygreen} $2877 \; \mathrm{cGy}$} & {\color{mygreen} $2094 \; \mathrm{cGy}$} \\
Anterior left eye & $\operatorname{D}_{0.1 \, \mathrm{cm}^3} \leq 500 \; \mathrm{cGy}$ & {\color{mygreen} $72 \; \mathrm{cGy}$} & {\color{mygreen} $62 \; \mathrm{cGy}$} \\
Anterior right eye & $\operatorname{D}_{0.1 \, \mathrm{cm}^3} \leq 500 \; \mathrm{cGy}$ & {\color{mygreen} $71 \; \mathrm{cGy}$} & {\color{mygreen} $61 \; \mathrm{cGy}$} \\
Posterior left eye & $\operatorname{D}_{0.1 \, \mathrm{cm}^3} \leq 500 \; \mathrm{cGy}$ & {\color{mygreen} $92 \; \mathrm{cGy}$} & {\color{mygreen} $79 \; \mathrm{cGy}$} \\
Posterior right eye & $\operatorname{D}_{0.1 \, \mathrm{cm}^3} \leq 500 \; \mathrm{cGy}$ & {\color{mygreen} $92 \; \mathrm{cGy}$} & {\color{mygreen} $83 \; \mathrm{cGy}$} \\
\end{tabular}
\end{center}
\egroup
}
\label{hnclingoalsresults}
\end{table}

\begin{figure}[h]
\centering
\begin{subfigure}[t]{0.34\textwidth}
\centering
\includegraphics[width = \textwidth]{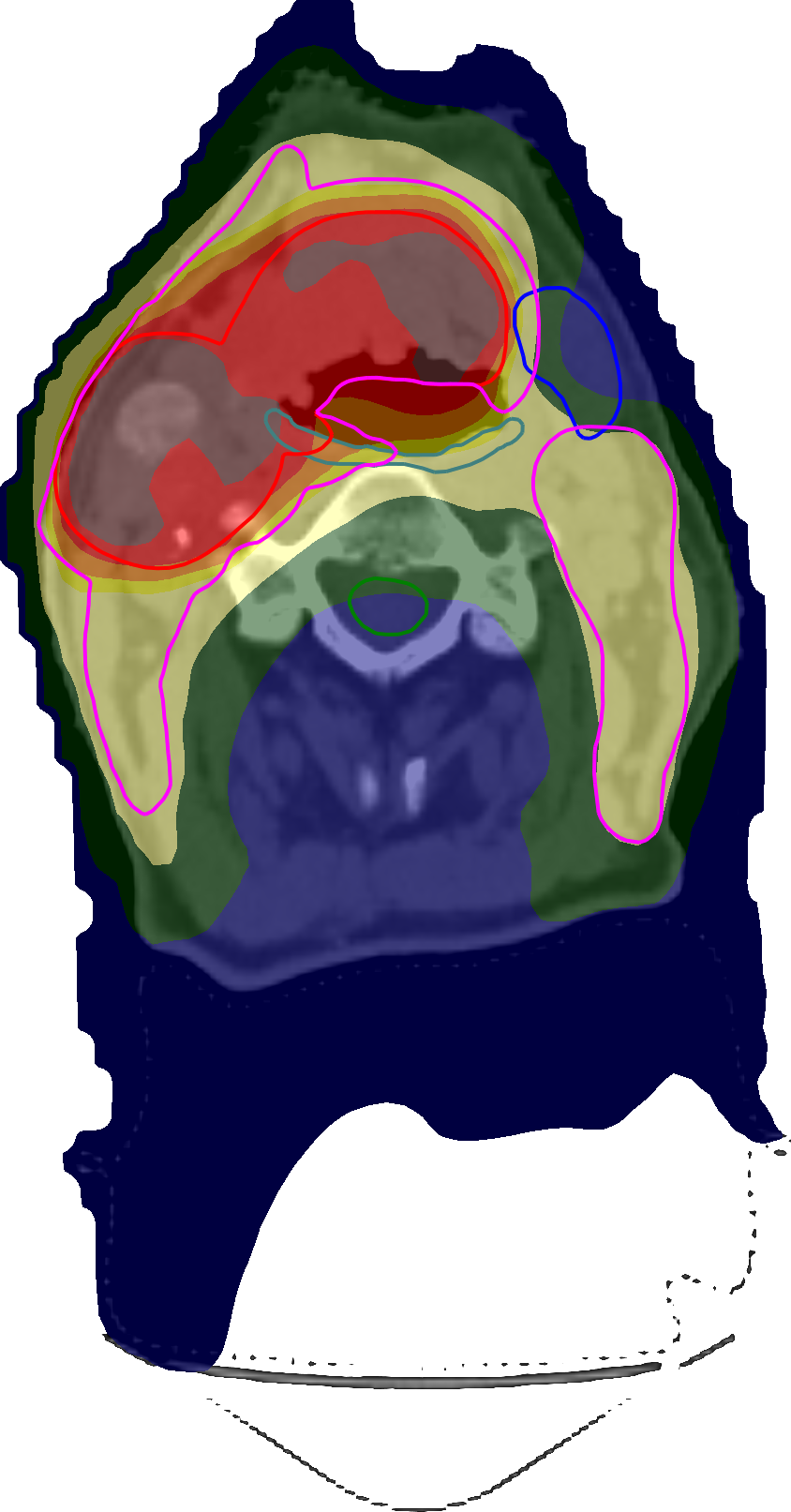}
\caption{}
\end{subfigure}%
\hspace{0.05\textwidth}
~ 
\begin{subfigure}[t]{0.46\textwidth}
\centering
\includegraphics[width = \textwidth]{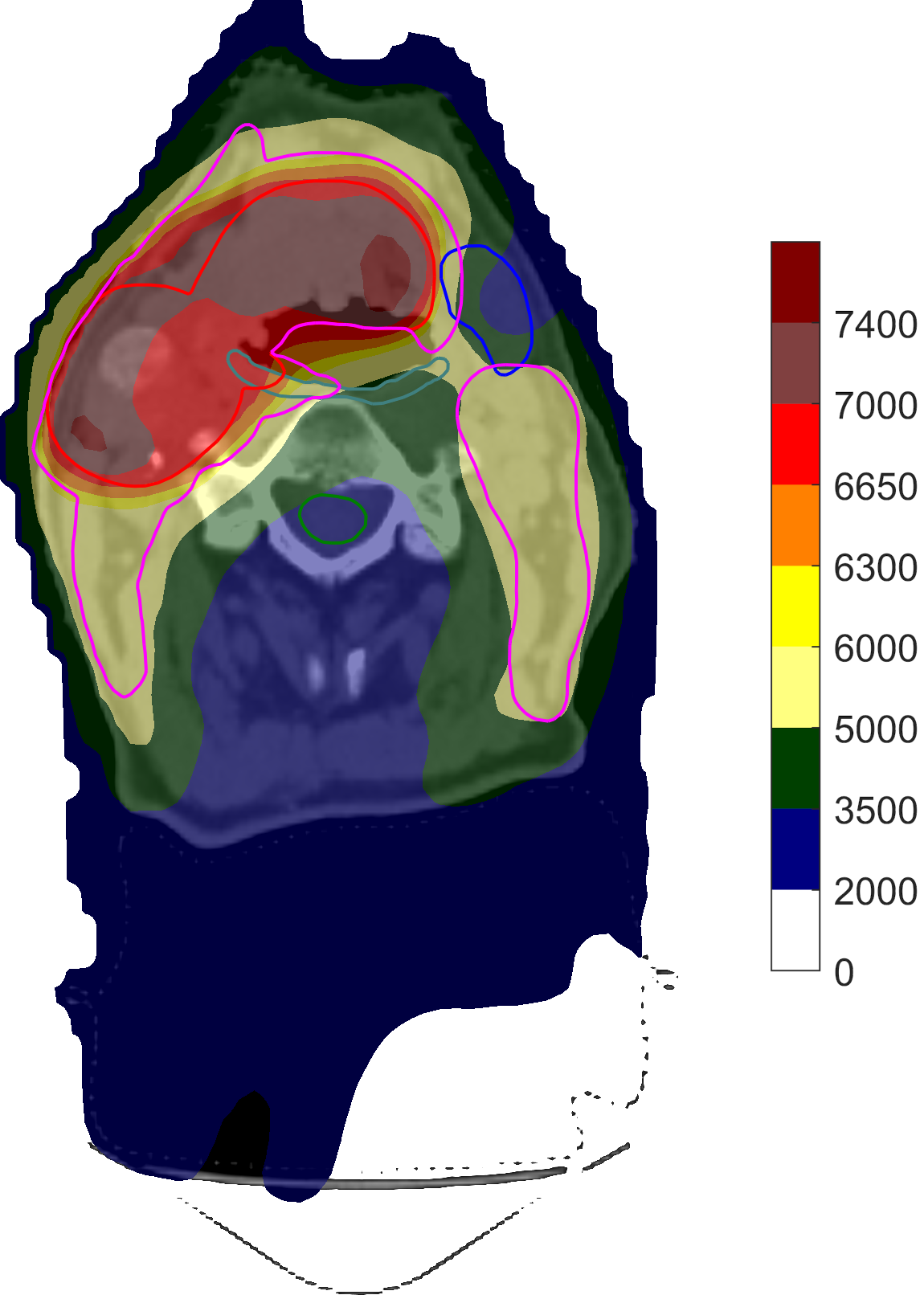}
\caption{}
\end{subfigure}\\[1ex]
\begin{subfigure}[t]{0.7\textwidth}
\centering
\includegraphics[width = \textwidth]{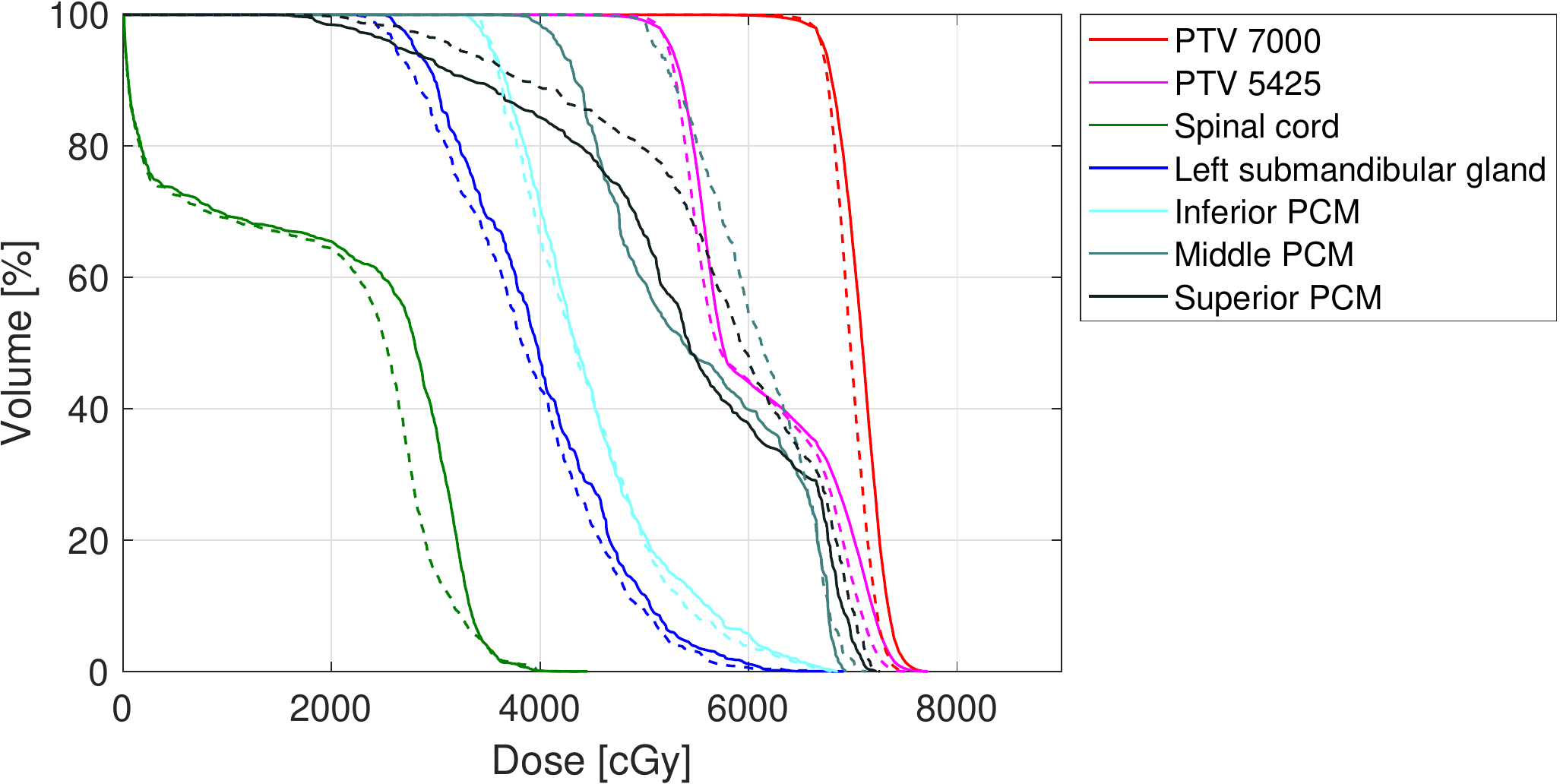}
\caption{}
\end{subfigure}
\caption{Transversal cuts of the spatial dose distributions of the optimized plans obtained for the constrained head-and-neck test using the conventional formulation (a) and using the direct optimization of clinical goals (b). A DVH comparison between the former (dashed) and the latter (solid) is shown in (c).}
\label{thirdhndvh}
\end{figure}

For the mostly constrained formulation, all optimizations started in the solution to the unconstrained formulation obtained from the direct optimization of clinical goals with mean-tail-dose, which was feasible with respect to all constraints. We chose the run using $\widetilde{w}_C^2 = 10^5$, which achieved the best overall plan quality of all values tried, to compare to the direct optimization---Table \ref{hnclingoalsresults2} shows the respective clinical goal levels after optimization and Figure \ref{thirdhndvh} shows the corresponding DVHs and spatial dose distributions. While the degrees of constraint infeasibility were similar, the direct optimization was able to lower the mean dose to the middle and superior PCMs significantly better than using the conventional functions. In particular, it was observed that the optimization using the conventional formulation quickly converged while that using the direct formulation was able to steadily push the mean doses down, again showcasing the problems with vanishing gradients associated with the conventional penalty functions.

\begin{table}[h]
\caption{Resulting clinical goal values for the mostly constrained head-and-neck case, compared between the conventional formulation and the direct optimization of clinical goals without mean-tail-dose. Not fulfilled goals are in red and fulfilled goals are in green.}
\vspace{-0.5cm}
{\bgroup
\def\arraystretch{1.15}
\begin{center}
\begin{tabular}{llll}
ROI & Goal & Conv & Direct 1 \\ \hline
PTV 7000 & $\operatorname{D}_{98 \, \%} \geq 6650 \; \mathrm{cGy}$ & {\color{myred} $6643 \; \mathrm{cGy}$} & {\color{myred} $6648 \; \mathrm{cGy}$} \\
PTV 7000 & $\operatorname{EUD}_1 \geq 6950 \; \mathrm{cGy}$ & {\color{mygreen} $6981 \; \mathrm{cGy}$} & {\color{mygreen} $7075 \; \mathrm{cGy}$} \\
PTV 7000 & $\operatorname{D}_{5 \, \%} \leq 7400 \; \mathrm{cGy}$ & {\color{mygreen} $7281 \; \mathrm{cGy}$} & {\color{mygreen} $7397 \; \mathrm{cGy}$} \\
PTV 5425 & $\operatorname{D}_{98 \, \%} \geq 5150 \; \mathrm{cGy}$ & {\color{myred} $5146 \; \mathrm{cGy}$} & {\color{mygreen} $5151 \; \mathrm{cGy}$} \\
$\text{PTV 5425} \setminus \text{PTV 7000}$ & $\operatorname{D}_{5 \, \%} \leq 5800 \; \mathrm{cGy}$ & {\color{mygreen} $5773 \; \mathrm{cGy}$} & {\color{myred} $5801 \; \mathrm{cGy}$} \\
External & $\operatorname{D}_{10 \, \%} \leq 3500 \; \mathrm{cGy}$ & {\color{mygreen} $2942 \; \mathrm{cGy}$} & {\color{mygreen} $3165 \; \mathrm{cGy}$} \\
Spinal cord & $\operatorname{D}_{0.1 \, \mathrm{cm}^3} \leq 4500 \; \mathrm{cGy}$ & {\color{mygreen} $3956 \; \mathrm{cGy}$} & {\color{mygreen} $3902 \; \mathrm{cGy}$} \\
Left parotid & $\operatorname{EUD}_1 \leq 2600 \; \mathrm{cGy}$ & {\color{mygreen} $2518 \; \mathrm{cGy}$} & {\color{mygreen} $2520 \; \mathrm{cGy}$} \\
Right parotid & $\operatorname{EUD}_1 \leq 2600 \; \mathrm{cGy}$ & {\color{mygreen} $2600 \; \mathrm{cGy}$} & {\color{myred} $2601 \; \mathrm{cGy}$} \\
Left submandibular gland & $\operatorname{EUD}_1 \leq 4000 \; \mathrm{cGy}$ & {\color{mygreen} $3867 \; \mathrm{cGy}$} & {\color{mygreen} $3997 \; \mathrm{cGy}$} \\
Brain & $\operatorname{D}_{0.1 \, \mathrm{cm}^3} \leq 5000 \; \mathrm{cGy}$ & {\color{mygreen} $2435 \; \mathrm{cGy}$} & {\color{mygreen} $2644 \; \mathrm{cGy}$} \\
Brainstem & $\operatorname{D}_{0.1 \, \mathrm{cm}^3} \leq 5600 \; \mathrm{cGy}$ & {\color{mygreen} $2056 \; \mathrm{cGy}$} & {\color{mygreen} $2160 \; \mathrm{cGy}$} \\
Anterior left eye & $\operatorname{D}_{0.1 \, \mathrm{cm}^3} \leq 500 \; \mathrm{cGy}$ & {\color{mygreen} $61 \; \mathrm{cGy}$} & {\color{mygreen} $63 \; \mathrm{cGy}$} \\
Anterior right eye & $\operatorname{D}_{0.1 \, \mathrm{cm}^3} \leq 500 \; \mathrm{cGy}$ & {\color{mygreen} $60 \; \mathrm{cGy}$} & {\color{mygreen} $61 \; \mathrm{cGy}$} \\
Posterior left eye & $\operatorname{D}_{0.1 \, \mathrm{cm}^3} \leq 500 \; \mathrm{cGy}$ & {\color{mygreen} $78 \; \mathrm{cGy}$} & {\color{mygreen} $79 \; \mathrm{cGy}$} \\
Posterior right eye & $\operatorname{D}_{0.1 \, \mathrm{cm}^3} \leq 500 \; \mathrm{cGy}$ & {\color{mygreen} $82 \; \mathrm{cGy}$} & {\color{mygreen} $82 \; \mathrm{cGy}$} \\
Inferior PCM & $\operatorname{EUD}_{1} \leq 5000 \; \mathrm{cGy}$ & {\color{mygreen} $4438 \; \mathrm{cGy}$} & {\color{mygreen} $4487 \; \mathrm{cGy}$} \\
Middle PCM & $\operatorname{EUD}_{1} \leq 5000 \; \mathrm{cGy}$ & {\color{myred} $6086 \; \mathrm{cGy}$} & {\color{myred} $5533 \; \mathrm{cGy}$} \\
Superior PCM & $\operatorname{EUD}_{1} \leq 5000 \; \mathrm{cGy}$ & {\color{myred} $5746 \; \mathrm{cGy}$} & {\color{myred} $5384 \; \mathrm{cGy}$} \\
\end{tabular}
\end{center}
\egroup
}
\label{hnclingoalsresults2}
\end{table}

\section{Discussion}
\label{discussion}

Treatment planning in radiation therapy often comprises several repetitions of optimizations with gradually adjusted parameters before a plan of clinically acceptable quality can be obtained, which may be a tedious process requiring continuous manual interaction. An essential cause of this is the distinction between optimization functions and the actual criteria used for evaluation of plan quality, arising from the disadvantageous mathematical properties of dose--volume criteria in their traditional definitions. 

Instead of resorting to surrogates such as the conventionally used quadratic penalty functions or trying to solve the mixed-integer programming formulation of optimization under dose--volume constraints, in this paper, we presented a new perspective of DVH-based metrics in general as functionals of a suitably defined random variable, formally establishing the equivalence to financial risk measures. By introducing the noise variable $\varepsilon$, we were able to obtain explicit formulas for gradients of smooth counterparts of common clinical goal functions such as volume-at-dose, dose-at-volume, mean-tail-dose, average dose, homogeneity index and conformity index. The result is a coherent framework in which one can use gradient-based solvers to optimize on any sufficiently regular function composed of DVH-based metrics, which makes it possible for the treatment planner to articulate \emph{a priori} preferences more accurately. As an example, we constructed a simple plan quality assessment metric consisting of weighted loss functions, but many more approaches are possible.

The numerical tests were performed on a prostate and a head-and-neck case using varying definitions of the plan quality metric. On the prostate case and the unconstrained head-and-neck case, it was shown that conventional penalty functions may come close to, but ultimately fail to, fully satisfy all clinical goals despite the fact that it is possible, as shown by the direct optimization of clinical goals. While most of the actual deviations in our cases were probably too small to be of any practical difference, it is not uncommon in other cases that even small infeasibilities in the clinical goals can render a plan clinically unacceptable. Regardless, the results highlight the fact that the proposed functions offer more precise control over clinical goal values. We also demonstrated that mean-tail-dose can provide for more effective reduction of extreme values in tail distributions, which often is desirable but not easily communicated through dose--volume criteria.

Although it is arguably easier to formulate a unified plan quality assessment metric by specifying loss functions of DVH-based metrics and their respective weights rather than weights of conventional penalty functions, the third test case, in which most clinical goals were set as constraints, served the purpose of reflecting a more traditional mindset---that is, to improve some objective goals as much as possible subject to fulfilling some constraint goals. Indeed, the only parameter to specify here was the constraint weight $w_C$, which is relatively easy to tune. While not addressing the question of how to fulfil the constraint goals in the first place (solving the previous unconstrained formulation is one alternative), this test showed that the proposed functions are significantly better than the conventional functions in pushing the objectives subject to a comparable degree of constraint conservativeness. Thus, although not completely eliminating the need for weights and fine-tuning, the proposed methods were shown to be powerful tools for reducing the need for time-consuming manual interaction.

The proposed framework for DVH-based metrics has the important advantage of requiring practically no parameter tuning---it is, according to our experience, not necessary to change the smoothness parameter $\epsilon$ between different patient cases. Also, approximation errors to conventional equivalents were found to be non-distinguishable in most cases. On a fundamental level, it is particularly attractive that conventional formulations and their respective smooth equivalents only differ through the value of $\epsilon$, facilitating the derivations of such equivalents for eventual other DVH-based or risk measure--inspired metrics.

As for mean-tail-dose, a downside of replacing dose-at-volume functions with mean-tail-dose functions at the same reference volume levels is that the clinical goals become more restrictive. Ideally, of course, one would like such clinical goals to be formulated in terms of mean-tail-dose from the beginning. Since they are not yet standard in clinical context, however, it would be beneficial to investigate how one can choose the reference volume level of the mean-tail-dose replacement in such a way that the resulting goal is somewhat equally restrictive as the original dose--volume goal. 

The discontinuity in the derivative of the linear-ramp loss functions led to some problems with clinical goals jumping in and out of fulfilment during the optimization. This is due to the loss becoming identically zero beyond fulfilment, which is inadequately taken advantage of by the solver. One can resolve this in future work by replacing the loss function by, for instance, soft-ramp functions \citep{albin} with the property that there is always incentive to improve upon the clinical goals. Also, since constraints defined directly in terms of clinical goals do not have the problem of gradients vanishing at the boundary of feasibility, as is the case with conventional penalty functions \citep{albin}, one would perhaps expect that optimization solvers may be less prone to violate such constraints. Experience has shown, however, that adding a constraint loss term $L_C$ in the total objective $L_{\mathrm{tot}}$ such as in Section \ref{optimizationformulation} tends to work better in practice, both in terms of computational time and resulting plan quality metric value. 

Apart from enabling the direct optimization of a given plan quality metric, the proposed framework for handling DVH metrics has applications in many different areas in treatment planning. One example is MCO, where a formulation as in \citet{lovisa} directly using dose--volume criteria (or mean-tail-dose criteria), rather than their corresponding quadratic penalty functions, becomes possible. The idea of using a constraint loss $L_C$ as described in Section \ref{optimizationformulation} may, moreover, improve the procedure of generating Pareto optimal plans by offering better control of nonlinear constraint infeasibilities. Another application of this idea is to various forms of lexicographic optimization \citep{miettinen}, where one tries to fulfil objectives in different levels of priority as an alternative to explicitly specifying a plan quality metric---this may, for example, be part of an automated planning algorithm or used to navigate automatically on a Pareto surface. Yet another application of direct clinical goal optimization is to various forms of dose mimicking, where the proposed functions can be used to ensure that certain clinical goals of importance are fulfilled in the reconstructed dose. 

\section{Conclusion}

In this paper, we presented a new perspective of DVH-based metrics as functionals of an auxiliary random variable, formalizing the often mentioned connection to risk measures in finance. By the alteration of a smoothness parameter, we obtained equivalents of common clinical goal functions such as volume-at-dose, dose-at-volume and mean-tail-dose as smooth functions and provided explicit expressions for their gradients, enabling the direct optimization of clinical goals. Numerical experiments performed on three test cases showed that this produced marginally better results in two of the cases and outperformed conventional penalty functions in the third case, judged by a pre-specified plan quality metric taking into account deviations in both objective and constraint goals---specifically, better pharyngeal constrictor sparing was achieved without sacrificing target coverage in the third case. Possible future work includes exploring other types of DVH-based metrics, loss functions and combinations thereof and investigating their advantages in other possible applications such as MCO and automated treatment planning. 

\begin{appendices}

\section{Proof of Proposition \ref{infdiffprop}}
\label{davinfdiff}

We show that $\mathrm{D}_v \in \mathcal{C}^{\infty}$ whenever $\epsilon > 0$ for all $0 < v < 1$, where $\mathcal{C}^n$ denotes the set of $n$ times continuously differentiable functions from $\mathbb{R}^{|R|}$ to $\mathbb{R}$. We proceed by induction and assume that $\mathrm{D}_v \in \mathcal{C}^n$ for some $n$ (the base case $n = 0$ is apparent). By Faà di Bruno's formula \citep{faa}, we can for any multi-index $\alpha = (\alpha_i)_{i \in R}$ with $|\alpha| = \sum_{i \in R} \alpha_i = n$ differentiate (\ref{dav}) to get
\begin{equation*}
\begin{split}
0 &= \sum_{i} r_{i} \sum_{\pi \in \Pi_n} K^{(|\pi|)}(d_{i} - \mathrm{D}_v(d)) \prod_{\beta \in \pi} \partial^{\beta} (d_{i} - \mathrm{D}_v(d)) \\
&= \sum_{i} r_{i} \sum_{\pi \in \Pi_n, |\pi| > 1} K^{(|\pi|)}(d_{i} - \mathrm{D}_v(d)) \prod_{\beta \in \pi} \partial^{\beta} (d_{i} - \mathrm{D}_v(d))  \\
&\quad\quad\quad + \sum_i r_i k(d_i - \mathrm{D}_v(d)) \, \partial^{\alpha} d_i - f_D(\mathrm{D}_v(d)) \, \partial^{\alpha}\mathrm{D}_v(d),
\end{split}
\end{equation*}
where $\partial^{\alpha} = \prod_{i \in R} \partial^{\alpha_i} / \partial d_i^{\alpha_i}$, where we have used $\Pi_n$ for the set of partitions of $\{1, \dots, n\}$ (each partition being a set of multi-indices), where $f_D$ is the density of $D$, and where $K^{(m)}$ denotes the $m$th derivative of $K$. Since the terms in the last display except the last one are $\mathcal{C}^1$ as they contain partial derivatives of $\mathrm{D}_v$ of at most order $n - 1$, and since $d \mapsto f_D(\mathrm{D}_v(d)) \in \mathcal{C}^n$ and $f_D(x) > 0$ for all $x$, we can rearrange to obtain $\partial^{\alpha} \mathrm{D}_v(d)$ as a ratio between a $\mathcal{C}^1$ function and a positive $\mathcal{C}^n$ function, which is again $\mathcal{C}^1$. Thus, we conclude that $\mathrm{D}_v \in \mathcal{C}^{n+1}$, which completes the induction step. The claim follows.

\section{Proof of Proposition \ref{closedformulasmtd}}
\label{mtdderivations}

We derive explicit formulas for function value and gradient of lower mean-tail-dose $\operatorname{MTD}_v^-$ and show that $-\operatorname{MTD}_v^-$ is a convex function of dose---the corresponding derivations for upper mean-tail-dose are analogous. Using the fact that $D$ has density $f_D(x) = \sum_{i \in R} r_i k(x - d_i)$, we have
\begin{equation*}
\begin{split}
\operatorname{MTD}_v^-(d) &= \frac{1}{1 - v} \sum_{i \in R} r_i \int_{-\infty}^{\mathrm{D}_v(d)} x k(x - d_i) \, dx \\ 
&= \frac{1}{1 - v} \sum_{i \in R} r_i \! \left( xK(x - d_i) \bigg|_{-\infty}^{\mathrm{D}_v(d)} - \int_{-\infty}^{\mathrm{D}_v(d)} K(x - d_i) \, dx \! \right) \\
&= \frac{1}{1 - v} \sum_{i \in R} r_i \Bigg( xK(x - d_i) \bigg|_{-\infty}^{\mathrm{D}_v(d)} \\
&\quad\quad\quad - \left( (x - d_i)K(x - d_i) + \epsilon^2 k(x - d_i) \right) \bigg|_{-\infty}^{\mathrm{D}_v(d)} \Bigg) \\
&= \frac{1}{1 - v} \sum_{i \in R} r_i \Big( d_i K(\mathrm{D}_v(d) - d_i) - \epsilon^2 k(\mathrm{D}_v(d) - d_i) \Big),
\end{split}
\end{equation*}
using the facts that $\lim_{x \to -\infty} xK(x - d_i) = 0$ and that $\partial(x K(x) + \epsilon^2 k(x)) / \partial x = K(x)$. Moreover, since $\partial k(x) / \partial x = -(x / \epsilon^2)k(x)$, we have for each $i \in R$ that
\begin{equation*}
\begin{split}
\frac{\partial \mathrm{MTD}_v^-(d)}{\partial d_i} 
&= \frac{1}{1 - v} \sum_{i' \in R} r_{i'} \bigg( \delta_{ii'} K(\mathrm{D}_v(d) - d_{i'}) \\
&\quad\quad\quad + (d_{i'} + \mathrm{D}_v(d) - d_{i'}) k(\mathrm{D}_v(d) - d_{i'}) \!\left( \frac{\partial \mathrm{D}_v(d)}{\partial d_i} - \delta_{ii'} \right) \!\! \bigg) \\
&= \frac{1}{1 - v} \bigg( r_i K(\mathrm{D}_v(d) - d_i) \\
&\quad\quad\quad + \mathrm{D}_v(d) \bigg( \frac{\partial \mathrm{D}_v(d)}{\partial d_i} f_D(\mathrm{D}_v(d)) - r_i k(\mathrm{D}_v(d) - d_i) \bigg) \! \bigg) \\
&= \frac{1}{1 - v} r_i K(\mathrm{D}_v(d) - d_i),
\end{split}
\end{equation*}
where the last equality is due to Proposition \ref{davgradprop}. 

To show that $-\operatorname{MTD}_v^-(d)$ is convex in $d$, it is sufficient to note the linearity of $D$ in $d$ for fixed outcomes of $I$ and $\varepsilon$ and apply Theorem 2 in \citet{rockafellaruryasev}. However, we provide an alternative proof here for the sake of instructiveness. Differentiating again, we get
\begin{equation*}
\begin{split}
\frac{\partial^2 \mathrm{MTD}_v^-(d)}{\partial d_i \, \partial d_{i'}} &= \frac{1}{1 - v} r_i k(\mathrm{D}_v(d) - d_i) \!\left( \frac{\partial \mathrm{D}_v(d)}{\partial d_{i'}} - \delta_{ii'} \right) \\
&= \frac{1}{(1 - v)f_D(\mathrm{D}_v(d))} \Big( r_i r_{i'} k(\mathrm{D}_v(d) - d_i) k(\mathrm{D}_v(d) - d_{i'}) \\
&\quad\quad\quad - \delta_{ii'} r_i k(\mathrm{D}_v(d) - d_i) f_D(\mathrm{D}_v(d)) \Big)
\end{split}
\end{equation*}
so that the Hessian $\partial^2 \mathrm{MTD}_v^-(d) / \partial d^2$ can be written as
\[
\frac{\partial^2 \mathrm{MTD}_v^-(d)}{\partial d^2} = \frac{1}{(1 - v)1^{\mathrm{T}} \kappa} \left( \kappa \kappa^{\mathrm{T}} - 1^{\mathrm{T}} \kappa \operatorname{diag} \kappa \right)\!
,\]
using the notation $1 = (1)_{i \in R}$ and $\kappa = (r_i k(\mathrm{D}_v(d) - d_i))_{i \in R}$. For the matrix inside the parentheses, we can for every $\xi \in \mathbb{R}^{|R|}$ write
\begin{equation*}
\begin{split}
\xi^{\mathrm{T}} \! \left( \kappa \kappa^{\mathrm{T}} - 1^{\mathrm{T}} \kappa \operatorname{diag} \kappa \right) \! \xi &= \big( \xi^{\mathrm{T}} \kappa \big)^2 - 1^{\mathrm{T}} \kappa \xi^{\mathrm{T}} (\operatorname{diag} \kappa) \xi \\
&= \big( \xi^{\mathrm{T}} (\operatorname{diag} \kappa) 1 \big)^2 - 1^{\mathrm{T}} (\operatorname{diag} \kappa) 1 \xi^{\mathrm{T}} (\operatorname{diag} \kappa) \xi \\
&= |\langle \xi, 1 \rangle|^2 - \Vert 1 \Vert \Vert \xi \Vert \\
&\leq 0,
\end{split}
\end{equation*}
where the last inequality is due to Cauchy--Schwarz with inner product and norm given by $\langle \xi, \xi' \rangle = \xi^{\mathrm{T}} (\operatorname{diag} \kappa) \xi'$ and $\Vert \xi \Vert = \sqrt{\langle \xi, \xi \rangle}$, which are well-defined since each component of $\kappa$ is positive. This shows that the Hessian of $\operatorname{MTD}_v^-(d)$ is negative semidefinite and thus that $-\operatorname{MTD}_v^-(d)$ is a convex function of $d$.

\end{appendices}

\printbibliography

\end{document}